%% file: X_in_C70.tex
\documentclass[aip,amsmath,amssymv, reprint]{revtex4-2}

\usepackage{graphicx}% Include figure files
\usepackage{float}
\usepackage{subfig}
\graphicspath{{./figures/}}
\usepackage{dcolumn}% Align table columns on decimal point
\usepackage{bm}% bold math
\usepackage[utf8]{inputenc}
\usepackage[T1]{fontenc}
\usepackage{mathptmx}
\usepackage{etoolbox}

\usepackage[version=4]{mhchem}
\usepackage{braket}
\usepackage{multirow} % Merge rows in table

\usepackage{orcidlink}
\usepackage{todonotes}
\newcommand{\Endo}[2]{\ce{#1}@\ce{#2}}
\newcommand{\wavenumber}{cm$^{-1}$ }

\makeatletter
\def\@email#1#2{%
 \endgroup
 \patchcmd{\titleblock@produce}
  {\frontmatter@RRAPformat}
  {\frontmatter@RRAPformat{\produce@RRAP{*#1\href{mailto:#2}{#2}}}\frontmatter@RRAPformat}
  {}{}
}%
\makeatother

\begin{document}
\title[\Endo{X}{C70}]{Exploring the parameter space of an endohedral atom in a cylindrical cavity}
% Force line breaks with \\
\author{K. Panchagnula*\, \orcidlink{0009-0004-3952-073X}}
\email{ksp31@cam.ac.uk}
 \affiliation{Yusuf Hamied Department of Chemistry, University of Cambridge, U.K.}%Lines break automatically or can be forced with \\
\author{A.J.W. Thom\,\orcidlink{0000-0002-2417-7869}}%
% \email{ajwt3@cam.ac.uk}

%\affiliation{ 
%Yusuf Hamied Department of Chemistry, University of Cambridge%\\This line break forced with \textbackslash\textbackslash
%}%

\date{\today}% It is always \today, today,

\begin{abstract}
	Endohedral fullerenes, or endofullerenes, are chemical systems of fullerene cages encapsulating single atoms or small molecules. These species provide an interesting challenge of Potential Energy Surface (PES) determination as examples of non-covalently bonded, bound systems. While the majority of studies focus on \ce{C60} as the encapsulating cage, introducing some anisotropy by using a different fullerene, e.g., \ce{C70} can unveil a double well potential along the unique axis. By approximating the potential as a pairwise Lennard-Jones (LJ) summation over the fixed \ce{C} cage atoms, the parameter space of the Hamiltonian includes three tunable variables: $(M,\varepsilon,\sigma)$ representing the mass of the trapped species, the LJ energy, and length scales respectively. Fixing the mass and allowing the others to vary can imitate the potentials of endohedral species trapped in more elongated fullerenes. We choose to explore the LJ parameter space of an endohedral atom in \ce{C70} with  $\varepsilon\in$ [20\wavenumber, 150\wavenumber], and $\sigma\in$ [2.85\AA , 3.05\AA]. 
	
	As the barrier height and positions of these wells vary between [1\wavenumber, 264\wavenumber] and [0.35\AA, 0.85\AA] respectively, using a 3D direct product basis of 1D harmonic oscillator (HO) wavefunctions centred at the origin where there is a local maximum is unphysical. Instead we propose the use of a non-orthogonal basis set, using 1D HO wavefunctions centred in each minimum and compare this to other choices. The ground state energy of the \Endo{X}{C70} is tracked across the LJ parameter space, along with its corresponding nuclear translational wavefunctions. A classification of the wavefunction characteristics, namely the prolateness and ``peanut-likeness'' based on its statistical moments is also proposed. Excited states of longer fullerenes are assigned quantum numbers, and the fundamental transitions of \Endo{Ne}{C70} are tracked across the parameter space.
\end{abstract}

\maketitle

\section{Introduction \label{sec:Intro}}
	\input{Introduction}
\section{Theory \label{sec:Theory}}
	\input{Theory}
\section{Results \label{sec:Results}}
	\input{Results}
\section{Conclusion \label{sec:Conc}}
	\input{Conclusion}

\section*{Supplementary Material}
	See the supplementary material for fullerene geometries.

\begin{acknowledgements}
	We wish to thank Malcolm H. Levitt, Richard J. Whitby and George R. Bacanu for their helpful comments on the manuscript and the experimental applications of this work. A.J.W.T and K.P. are thankful to the Walters Kundert Next Generation Fund for funding on this project.
\end{acknowledgements}

\section*{Data Availability Statement}
The data that support the findings of this study are openly available in Apollo - University of Cambridge Repository at https://doi.org/10.17863/CAM.100065\cite{KripaPanchagnulaDataset2023}, reference number \citenum{KripaPanchagnulaDataset2023}.

\bibliography{X_in_C70}
\end{document}

%% file: Introduction.tex
%\documentclass[./X_in_C70]{subfiles}
%\graphicspath{{\subfix{./figures/}}}
%\begin{document}
	Endohedral fullerenes or endofullerenes are molecular complexes where a fullerene cage traps an atom or molecule(s). The development of the technique known as ``molecular surgery''\cite{murataSynthesisReactionFullerene2008, bloodworthSynthesisEndohedralFullerenes2022}
	 has allowed for controlled synthesis of not just noble-gas or metallo-endofullerenes, but also light molecular endofullerenes including \Endo{H2}{C60}\cite{mamoneRotorCageInfrared2009}, \Endo{H2O}{C60}\cite{krachmalnicoffOptimisedScalableSynthesis2014} and \Endo{CH4}{C60}\cite{bloodworthFirstSynthesisCharacterization2019}.
	  The fullerene cage almost completely shields the guest species from the environment, and these isolated species epitomise behaviour dominated by quantum phenomena. 
	
	These strong quantum effects arise due to the encapsulating potential quantising the translational motion, and the coupling of these modes with with the rotational states of the trapped molecule\cite{bacicPerspectiveAccurateTreatment2018}. The spectroscopic fine structure associated with this translational-rotational coupling has given rise to numerous theoretical studies focusing on \Endo{H2}{C60}\cite{xuCoupledTranslationrotationEigenstates2009, yeSolid60Coupled2013, xuH2HDD22008, xuInelasticNeutronScattering2011, geInfraredSpectroscopyEndohedral2011, yildirimRotationalVibrationalDynamics2002, felkerTranslationrotationStates602016}
	 and \Endo{H2O}{C60}\cite{felkerCommunicationQuantumSixdimensional2016, felkerElectricdipolecoupledH2OC602017,felkerFlexibleWaterMolecule2020, rashedInteractionsWaterMolecule2019, felkerExplainingSymmetryBreaking2017,carrillo-bohorquezEncapsulationWaterMolecule2021, beduzQuantumRotationOrtho2012}.
	  In order to understand the coupling, the translational-rotational eigenstates must be found which requires construction of a Potential Energy Surface (PES). For both of these major systems, the PES is usually generated by assuming a pairwise-additive Lennard-Jones (LJ) form\cite{bacicPerspectiveAccurateTreatment2018, mandziukQuantumThreeDimensional1994}, with energy scale $\varepsilon$ and length scale $\sigma$, as this is well suited for modelling dispersion interactions\cite{bugNonlinearVibrationalDynamics1992}, with the parameters fitted to match the spectroscopic data\cite{xuCoupledTranslationrotationEigenstates2009}.
	   Calculating an accurate ab-initio surface is computationally intensive because of the large number of atoms and interactions dominated by dispersion effects, and so only the PES of \Endo{HF}{C60} has been acquired using these methods\cite{kaluginaPotentialEnergyDipole2017}.  
	
	While the majority of studies have focused on \ce{C60} as the encapsulating fullerene due to its high $I_h$ symmetry and wealth of experimental data, this is not the only available choice. The next most abundant fullerene, \ce{C70}, has its symmetry reduced by elongating the buckyball along one axis and its cavity can contain larger or more molecules including $\ce{(H2)_2}$@\ce{C70}\cite{sebastianelliHydrogenMoleculesFullerene2010} and  $\ce{(H2O)_2}$@\ce{C70}\cite{zhangSynthesisDistinctWater2016}.
	 Elongating the cage even further in a similar manner produces larger fullerenes or fullertubes in the $D_{5d/h}$ series which have been recently synthesised\cite{schusslbauerExploringThresholdFullerenes2022, koenigFullertubesCylindricalCarbon2020, liuGiganticC120Fullertubes2022, schiemenzVibrationalAnatomyC902022}.
	  An important feature which arises in the PES of these lower symmetry fullerenes/fullertubes is a symmetric double well potential along the unique axis. Traversing this series for the fixed LJ parameters of \Endo{Ne}{C70} given in Table \ref{table: Ne LJ parameters} towards larger cages pulls the minima further apart, and increases the barrier height. The first row of parameters have been used in the calculation of the endohedral vibrational levels \Endo{Ne}{C70}\cite{mandziukQuantumThreeDimensional1994, bugNonlinearVibrationalDynamics1992}, but the latter set were used in the calculation of the potential of noble gas endohedral and exohedral complexes with \ce{C60} and \ce{C70}\cite{pangEndohedralEnergiesTranslation1993, jimenez-vazquezEquilibriumConstantsNoble1996} and are more commonly used. 
	  
  	\begin{table}
		\begin{ruledtabular}
			\begin{tabular}{ccc}
				$\varepsilon$/\wavenumber&$\sigma$/\AA & Reference \\
				\hline 43.79&3.03 & Ref \citenum{mandziukQuantumThreeDimensional1994}\\
				26.57&3.029 & Ref \citenum{jimenez-vazquezEquilibriumConstantsNoble1996}
			\end{tabular}
		\end{ruledtabular}
		\caption{Lennard Jones parameters for \ce{Ne} in a fullerene.}
		\label{table: Ne LJ parameters}
	\end{table}
	
	\begin{figure}
		\includegraphics[scale=0.48]{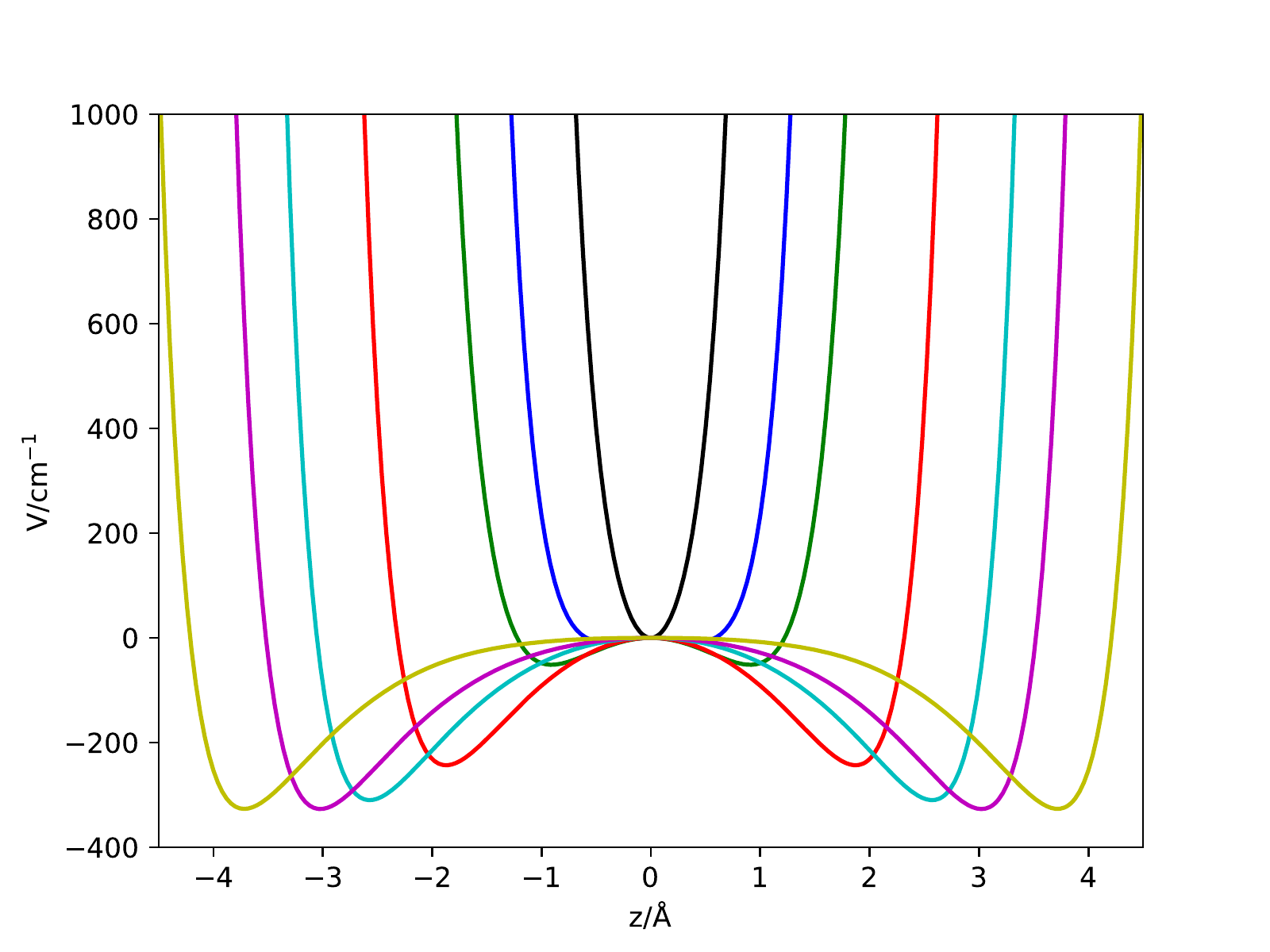}
		\caption{Pairwise summed LJ potential along the unique axis traversing the fullerene/fullertube $D_{5d/h}$ series for $(\varepsilon, \sigma)$=(43.79\wavenumber, 3.03\AA). \ce{C60} is barrier-less shown in black, \ce{C70} in blue, \ce{C80} in green, \ce{C90} in red, \ce{C100} in cyan, \ce{C110} in magenta and \ce{C120} in yellow.}
		\label{fig: Fullertube Double Well}
	\end{figure}
	
	In order to garner some intuition of the behaviour of a species trapped in a cylindrical cavity created by a fullerene from this series, knowledge of the PES is required. While either the endohedral species or the encapsulating fullerene can be altered, the limited choice in cages is restrictive. The alternative we choose to adopt is to fix the fullerene cage and vary both LJ parameters in order to more flexibly change the nature of the double well. This allows for more careful tuning of the barrier height and positions of the minima, similar to what is seen when elongating the fullerene along the $D_{5d/h}$ series. The geometries of the fullerenes \ce{C60}, \ce{C70}, \ce{C80}, \ce{C90}, \ce{C100}, \ce{C110}, and \ce{C120}\cite{liuGiganticC120Fullertubes2022, bakerStructurePropertiesC701991, schwerdtfegerProgramFullereneSoftware2013} compared in Figure \ref{fig: Fullertube Double Well} were oriented to make the unique axis the $z$ direction and symmetrised using IQMol \cite{gilbertIQmol32023}, and their symmetries validated\cite{beruskiAlgorithmsComputerDetection2014, BangHuynhPolyinspect2023} using in-house Python software. Adding an extra belt of \ce{C} atoms pulls the minima further apart, but also increases the barrier height. Given the form of the potential, these quantities are quite sensitive to the size of the cage, and so changing the \ce{C-C} bond length can also alter them. Changing the LJ parameters (even though the mass is fixed) can also correspond to changing the endohedral species. The LJ parameters for both the noble gas\cite{foroutanAdsorptionBehaviorTernary2010, pangEndohedralEnergiesTranslation1993, jimenez-vazquezEquilibriumConstantsNoble1996, albertSimulationsXeC602008}
	 and metal\cite{dhimanOptimizedInteractionParameters2017, verkhovtsevSoftLandingMetal2020, neek-amalFormationAtomicNanoclusters2009}
	 -fullerene interactions can take differing values depending on the environment where they were calculated.
	
	The specific LJ parameters of the endohedral complex can be found by either considering the atomic polarisabilities and susceptibilites\cite{bugNonlinearVibrationalDynamics1992}, or by minimising differences to experimental data\cite{xuCoupledTranslationrotationEigenstates2009}.

	When the parameters are found by minimising the difference to the experimental data, there can be multiple solutions for them	 which further justifies the choice of allowing the LJ parameters to take a range of values.
	  Varying the LJ parameters instead of just the fullerene cage allows the barrier height $B$ and minima location $z_{\text{min}}$ to change smoothly. This provides more control in tuning these quantities, along with the zero-point energy ($ZPE$) as its relation to $B$ partitions the parameter space into two main regions: $B>ZPE$ and $B<ZPE$. 
	 
	In the region where $B<ZPE$, which is the case for \Endo{Ne}{C70}\cite{mandziukQuantumThreeDimensional1994},
	 the atom can ``tunnel'' between both minima and the simple harmonic oscillator (HO) basis set used within the discrete variable representation (DVR) calculations is perfectly suited to this region. However, as the minima get pulled apart, and as the barrier height increases as seen in Fig \ref{fig: Double Well}, using these functions centred at the origin in the finite basis representation (FBR) becomes unphysical as the potential is a local maximum. Instead of this we propose and benchmark an alternative basis with HO functions centred at both minima. Although this basis is non-orthogonal, the added flexibility can recover results in the usual case of $B<ZPE$, and we show it is much better suited when the alternative scenario of $B>ZPE$ arises. These two regimes will lead to different forms of ground state wavefunction: one that has a single maximum at the centre of the cage, or one that exhibits two maxima above the minima in the potential. This classification is discrete and assignment is not necessarily obvious, hence a system based on the statistical moments of the wavefunction density is proposed. This description is continuous and may correspond to some experimental observables. 	
	 
	 Alongside information about the ground state of the system, diagonalising the Hamiltonian also converges the energies of excited states which can be used to calculate transition energies. These can be discerned spectroscopically, and when compared to the values calculated by this technique, the LJ parameters that best describe the \ce{X-C} interaction can be ascertained.
	 
	In this paper we consider a generic system of  \Endo{X}{C70} where \ce{X} represents a single particle, e.g. a noble gas atom or the centre of mass of a small molecular species, and vary the LJ parameters to explore a range of barrier heights and minima locations which imitate what could be encountered with a fullerene further down the $D_{5d/h}$ series. The behaviour of the ground state is tracked across a region of this parameter space, and the differing shapes of the translational wavefunction are classified. Fundamental transition energies of \Endo{X}{C70} are also calculated across the LJ parameter space, alongside excited state energies of \ce{X}@\ce{C}$_{\mathrm{n}}$, where $n\in[70,80,90,100]$ for a fixed set of LJ parameters. The relevant background theory on the diagonalisation of the Hamiltonian and wavefunction analysis is discussed in Section \ref{sec:Theory}, with the results and discussion in Section \ref{sec:Results}. Section \ref{sec:Conc} contains the conclusions drawn from the systems studied and outlines possible future work, both theoretical and experimental.
%\end{document}

%% file: Theory.tex
%\documentclass[./X_in_C70]{subfiles}
%\graphicspath{{\subfix{./figures/}}}
%\begin{document}
	\subsection{Choosing a Translational Basis Set \label{sec:Basis Set}}

		Using the 6-12 pairwise LJ potential, the Hamiltonian for \Endo{X}{C70} in atomic units can be written as 
			\begin{equation}
				\hat{H}=-\frac{1}{2M}\nabla^2 + 4\varepsilon\sum_i\left[\left(\frac{\sigma}{r_i}\right)^{12}-\left(\frac{\sigma}{r_i}\right)^{6}\right],\label{eq:LJ Hamiltonian}
			\end{equation}
		where $M$ is the effective two-body reduced mass of the endohedral complex, $(\varepsilon,\sigma)$ are the energy and length scale of the \ce{X-C} interaction\cite{mandziukQuantumThreeDimensional1994}.
		 The sum is taken over all fixed positions of the carbon atoms, with $r_i$ being the distance between the endohedral atom and the $i$th carbon atom. Due to the anisotropy in the cage, this is treated using a 3D finite basis representation (FBR) constructed as a direct product of 1D HO wavefunctions expanded from the origin
			\begin{align}
				\ket{n_r}&=\frac{1}{\sqrt{2^nn!\sqrt{\pi}}}H_n(q_r)\exp(-\frac{1}{2}q_r^2) \label{eq: 1D HO}\\
				\ket{n_xn_yn_z}&=\ket{n_x}\ket{n_y}\ket{n_z} \label{eq: 3D HO Basis}
			\end{align}

		where $q_r=\sqrt{\alpha}r$ is a scaled dimensionless coordinate, and $r$ represents a single Cartesian coordinate. Instead of working in the FBR, the Hamiltonian can be diagonalised within a DVR\cite{lightGeneralizedDiscreteVariable1985}
		 which requires choosing both an underlying basis set, as well as the set of points where these functions are centred. Choosing the same basis functions for the DVR, this is isomorphic to the FBR as they can be interconverted using a unitary transformation, with the advantage being that the potential energy matrix in the DVR is diagonal. Traditionally, the DVR points are chosen by selecting an interval for each coordinate alongside an energy cutoff and only the points where the potential lies under this are kept. Alternatively, a Potential-Optimised DVR\cite{echavePotentialOptimizedDiscrete1992}
		  (PODVR) scheme can be considered where the scaling factor for the dimensionless coordinate $q_r$ is determined from the potential. This can be chosen by either taking the second derivative of the potential at the minimum and using the relation $\alpha = \sqrt{k_{\text{eff}}M}$, or by leveraging the variational principle and using the ground state harmonic oscillator as a trial wavefunction in 1D %cite DVR, PODVR
			\begin{equation}
				E(\alpha)=\frac{\Braket{\exp\left(-\frac{1}{2}\alpha r^2\right)|\hat{H}|\exp\left(-\frac{1}{2}\alpha r^2\right)}}{\sqrt{\pi/\alpha}}\label{eq: Variational Principle}.
			\end{equation}
		By minimising $E(\alpha)$, a more appropriate effective force constant is recovered which corresponds to a higher zeroth order frequency. This is because the procedure accounts for the higher degree nature of the potential. This strategy is required for the unique axis even when $B$ is small as the origin is a local maximum which would result in unbound basis states.
		
	\subsection{Tackling the Double Well \label{sec:Double Well}}
	The \ce{C70} cage is oriented such that the $z$-axis is the unique axis, with the $x$-axis aligned to be one of the $C_2'$ rotation axes. As $B$ rises, the aforementioned PODVR method starts to break down as the depth of the well weighs down the effective parabola decreasing $\alpha$. This necessitates the use of more basis functions (or equivalently, quadrature points), more of which are placed towards the edge of the cage, and even outside it. The potential in this region is of the order of 1,000,000 \wavenumber and this leads to unstable solutions. This can be tackled in two different ways, either from a DVR perspective or from the FBR.
		\begin{figure}
			\subfloat[$yz$]{\includegraphics[scale=0.24]{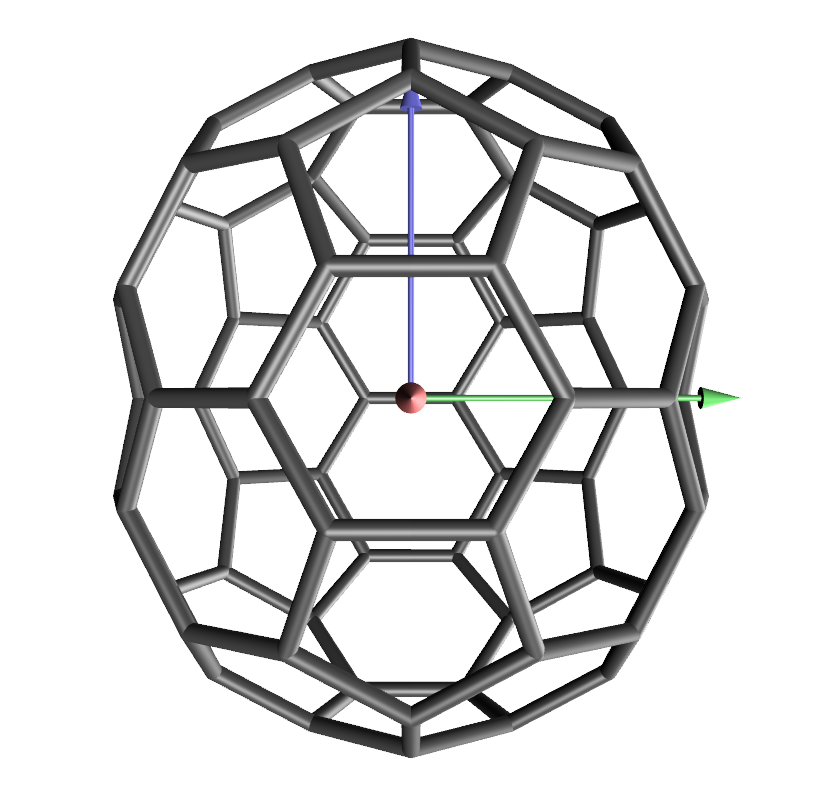}}\,
			\subfloat[$xy$]{\includegraphics[scale=0.24]{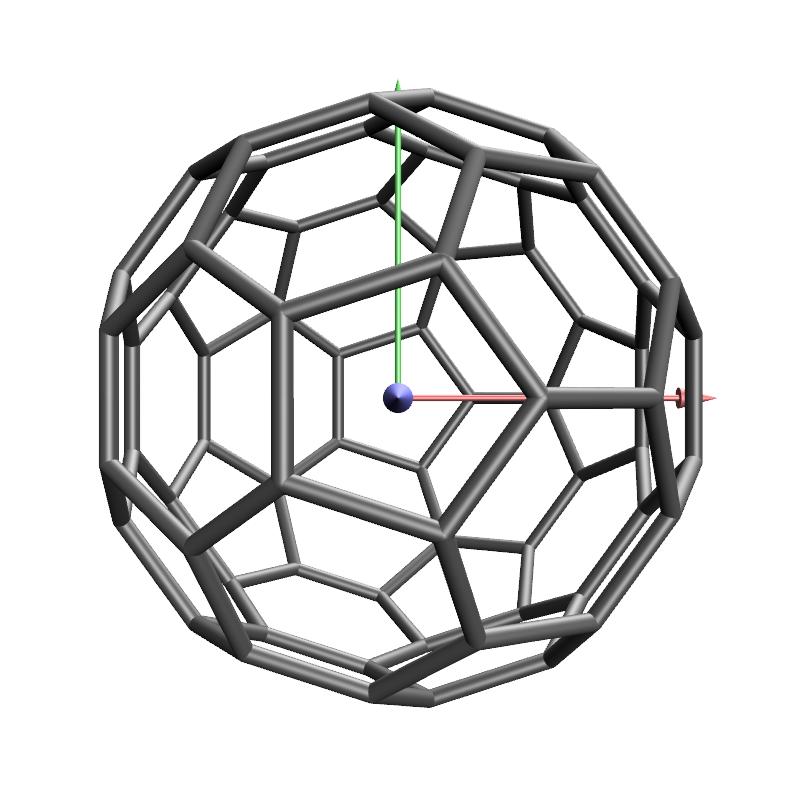}}
			\caption{Orientation of the fixed \ce{C70} cage in the (a) $yz$ and (b) $xy$ planes. The red, green and blue arrows correspond to the positive $x$, $y$, and $z$ Cartesian directions respectively.}
			\label{fig: C70}
		\end{figure}
		
	From the DVR viewpoint, accounting for the origin being a local maximum, a possible choice of basis set is to use sinc functions \cite{colbertNovelDiscreteVariable1992}.
	 The advantage is that this is an orthogonal basis set, and the kinetic energy matrix elements are known analytically. However, these functions take no account of the nature of the potential and have equidistant sampling points. The further the minima are located from the origin, or as $B$ rises, the more quadrature points that will be needed. An alternative, ignoring the fact that the origin is a local maximum, is to revert to the traditional DVR standpoint and impose the range of sampling points, instead of optimising the scaling parameter. Using the 1D HO basis functions as in Equation \eqref{eq: 1D HO}, this dictates an effective harmonic potential and fixes the scaling parameter. This maintains the traditional advantages of the DVR, but (ignoring truncation effects) the double well may imply that the energy doesn't always decrease with increasing number of basis functions.
	 
	 From the FBR perspective, both minima in the double well can be acknowledged with basis functions expanded around them instead of the local maximum origin. Using 1D HO wavefunctions along the unique direction, the scaled dimensionless coordinate and basis functions can be written as
			\begin{align}
				q^{\pm}_z&=(z\pm a)\sqrt{\alpha} = q_z\pm a_z\label{eq: 1D HO Scale}\\
                \ket{n^{\pm}_z}&=\frac{1}{\sqrt{2^nn!\sqrt{\pi}}}H_{n}(q^{\pm}_z)\exp{-\frac{1}{2}(q^{\pm}_z)^2}\label{eq: 1D HO Shifted},
			\end{align}
	where $a$ is the location of the minimum of the potential. A similar optimisation choice for $\alpha$ as in Section \ref{sec:Basis Set} exists, but here it is calculated by taking the second derivative at the minimum. As these functions are confined in a single minimum, generating delocalised wavefunctions that span both minima is achieved by taking in-phase and out-of-phase combinations of equivalent basis functions
			\begin{equation}
				\ket{n_z} = \frac{1}{\sqrt{2}}\left(\Ket{\left\lfloor\frac{n_z}{2}\right\rfloor^-}+(-1)^{\left\lfloor\frac{n_z}{2}\right\rfloor + (n_z\text{ mod 2})}\Ket{\left\lfloor\frac{n_z}{2}\right\rfloor^+}\right) \label{eq: Double Min Basis}
			\end{equation}
		where the kets are defined as in Equation \eqref{eq: 1D HO Shifted} and the phase factor chosen to maintain the expected nodal structure with increasing quantum number.  This symmetrised double minimum basis set is somewhat reminiscent of the orbitals and configuration basis used when considering the interaction of two \ce{H2} molecules within \ce{C70}\cite{felkerNuclearorbitalConfigurationinteractionStudy2013}. As this basis is adapted from the potential, fewer points are expected to be needed to converge the eigenstates. 
			\begin{figure}
				\includegraphics[scale=0.45]{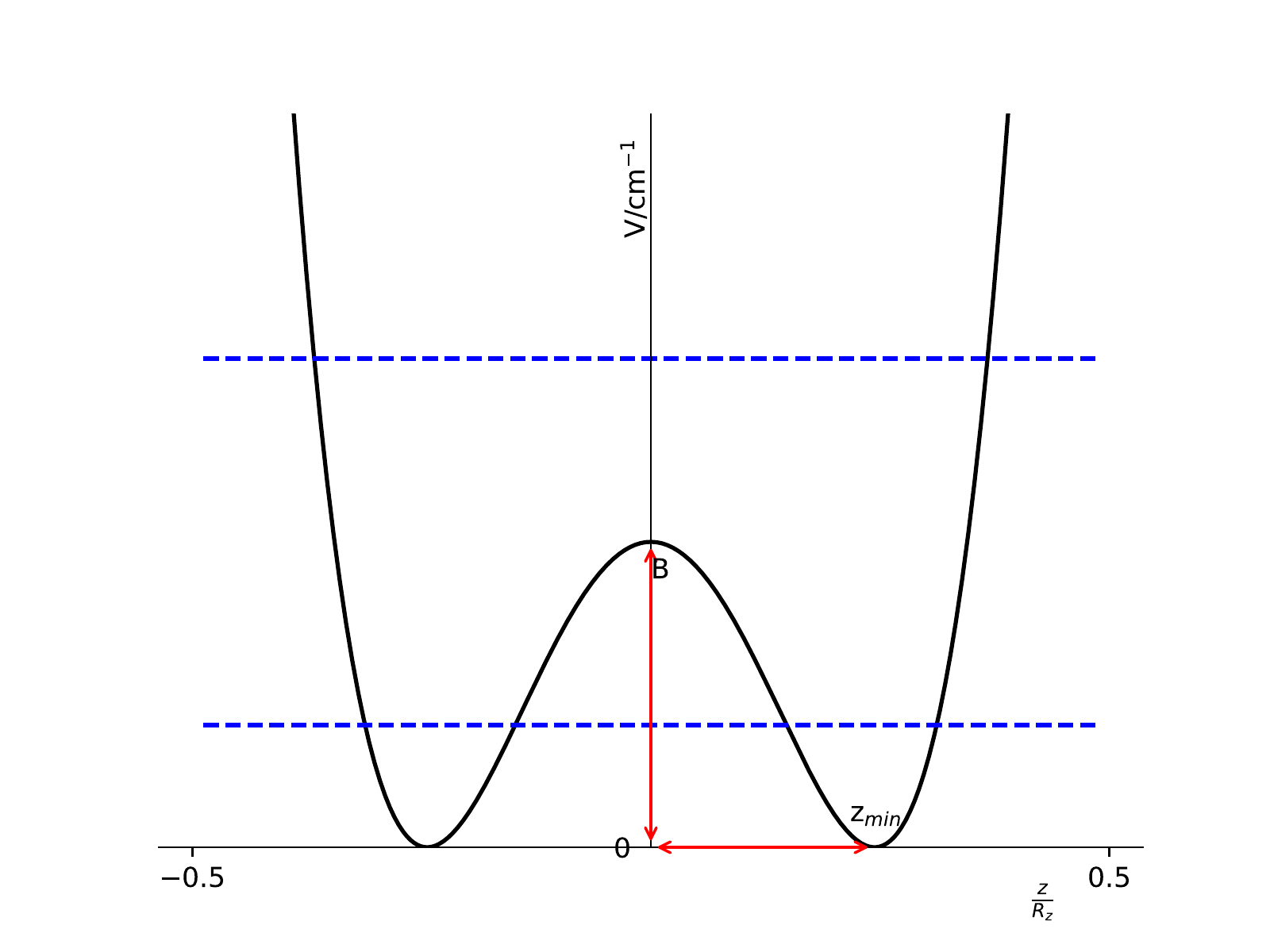}
				\caption{A generic symmetric double well potential with barrier height $B$ and $z_{\text{min}}$ indicated with red arrows. The dashed blue lines indicate two different regions of $ZPE$, smaller or larger than $B$.}
				\label{fig: Double Well}
			\end{figure}					
		
		On the other hand the Hamiltonian cannot be simply manipulated such that these functions form an eigenbasis and so matrix elements have to be calculated explicitly. This can be done using Gauss-Hermite quadrature, further justifying this choice of basis set. The Hamiltonian in Equation \eqref{eq:3D Hamiltonian} is split up into four terms: $\hat{h}^0_x$ with eigenfunctions $\ket{n_x}$ and frequency $\omega_x$, $\hat{h}^0_y$ with eigenfunctions $\ket{n_y}$ and frequency $\omega_y$. These chosen to be 1D HO operators, frequencies and wavefunctions of the form given in Equation \eqref{eq: 1D HO}. $\hat{k}_z$ is the kinetic energy operator along the $z$ direction and $\Delta V$ is the potential not accounted for in the definitions of $\hat{h}^0_x$ and $\hat{h}^0_y$. The Hamiltonian and overlap matrix elements have the following forms

			\begin{align}
				\hat{H}&= \hat{h}^0_x+\hat{h}^0_y+\hat{k}_z+\Delta V \label{eq:3D Hamiltonian}\\
				\braket{m_xm_ym_z|\hat{H}|n_xn_yn_z}&=(n_x+\frac{1}{2})\omega_x\delta_{n_xm_x}\delta_{n_ym_y}\braket{m_z|n_z}\nonumber\\
				&+(n_y+\frac{1}{2})\omega_y\delta_{n_xm_x}\delta_{n_ym_y}\braket{m_z|n_z}\nonumber \\
				&+\delta_{n_xm_x}\delta_{n_ym_y}\braket{m_z|\hat{k}_z|n_z} \nonumber\\
				&+\braket{m_xm_ym_z|\Delta V|n_xn_yn_z}\label{eq: Hmn}\\
				\braket{m_xm_ym_z|n_xn_yn_z}&=\delta_{n_xm_x}\delta_{n_ym_y}\braket{m_z|n_z}\label{eq: Smn}.
			\end{align}

	Depending on the separation of the two minima, the overlap integral $\braket{m^-|n^+}$ could be non-zero making the basis set non-orthogonal.  Carelessly increasing the basis set size will lead to some linear dependence within the basis set. This overcompleteness arises when states with zeroth-order energies in the $z$ direction larger than B start being included within the basis. The energies are given by $(n_z+\frac{1}{2})\omega_z$, where $\omega_z$ is the frequency within the well. These functions are very diffuse and taking linear combinations of these eventually leads to overcompleteness. This is alleviated by projecting the Hamiltonian and overlap matrices into a linearly independent subspace before diagonalising.

	\subsection{Wavefunction Classification \label{sec:Wavefunc Stats}}

		In the regime $B<ZPE$, the ground state wavefunction is ellipsoidal. This shape can be described by the length of its semi-axes, but this requires selecting an isosurface to measure from. An alternative that can be calculated is the standard deviation in each direction. Leveraging the isotropy in the $x$ and $y$ directions, the prolateness of the spheroid, $\varsigma^2$ can be calculated by considering the ratio of variances in the $z$ and $x$ directions
			\begin{equation}
				\varsigma^2 = \frac{\sigma_z^2}{\sigma_x^2} = \frac{\braket{0|z^2|0}-\braket{0|z|0}^2}{\braket{0|x^2|0}-\braket{0|x|0}^2}= \frac{\braket{0|z^2|0}}{\braket{0|x^2|0}},\label{eq: Variance Ratio}
			\end{equation}
		
		where the last equality follows as $\mu_z\equiv\braket{0|z|0}$ and $\mu_x\equiv\braket{0|x|0}$ are zero by symmetry. As B and ZPE become more comparable in size, and eventually when $B>ZPE$, the wavefunction along the $z$ direction moves away from being singly-peaked at the origin to having two equivalent maxima. This moves the wavefunction away from an ellipsoid, to a more peanut-like shape. The simplest way to classify this would be to just consider $\sigma_z^2$, which would be large when there is more density away from the mean. Although this will be the case for a double-maxima wavefunction, it could be generated fictitiously with a shallow and very diffuse single peaked ground state. Instead,  the kurtosis of the normalised state can be calculated as
		\begin{equation}
			\kappa = \Braket{0|\left(\frac{z-\mu_z}{\sigma_z}\right)^4|0}=\frac{1}{\sigma_z^4}\braket{0|z^4|0}. \label{eq: Kurtosis}
		\end{equation}
		This accounts for both the mean and variance of the distribution along the unique direction and measures the density at $\mu_z\pm\sigma_z$. It can be thought of as the relative weight of the shoulder of the distribution compared to the mean and tails. For a pure Gaussian distribution, the kurtosis is strictly equal to 3. A bimodal distribution which corresponds to a double peaked wavefunction has a kurtosis in the interval [1,3). % citation on kurtosis?

		Together, these two statistics describe the deviation of the shape of the ground state from an idealised spherical wavefunction which is the shape for an atom inside \ce{C60}\cite{bacanuExperimentalDeterminationInteraction2021c}.
		 The variance ratio measures the ellipsoidal nature and the kurtosis gives a measure of the peanut-like shape. 
%\end{document}

%% file: Results.tex
%\documentclass[./X_in_C70]{subfiles}
%\graphicspath{{\subfix{./figures/}}}
%\begin{document}
	\subsection{Computational Details \label{sec:Comp}}
		Along the $x$ and $y$ directions, 12 basis functions of the form shown in Equation \eqref{eq: 1D HO} were used, with the scale factor determined by minimising Equation \eqref{eq: Variational Principle} with the corresponding one-dimensional Hamiltonian. For the unique direction, the choice between sinc DVR functions, HO DVR functions, or basis functions of the form shown in Equation \eqref{eq: Double Min Basis} was determined by the convergence of the ground state energy of a fixed Hamiltonian.

			\begin{figure}
				\subfloat[1D double well potential with LJ parameters (43.79\wavenumber, 3.03\AA). This corresponds to the region where $ZPE\approx B$.]{\includegraphics[scale=0.48]{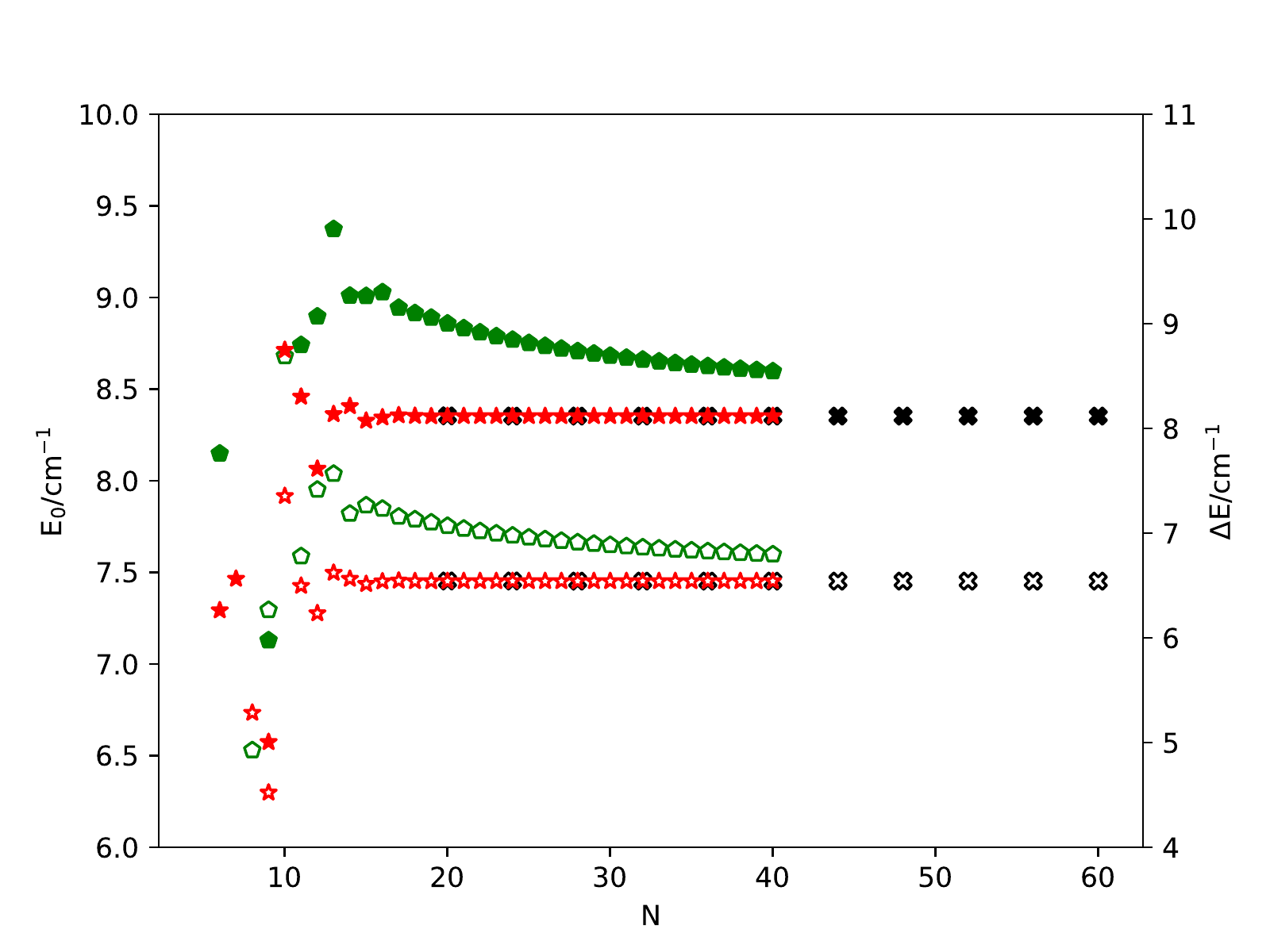}}\\
				\subfloat[1D double well potential with LJ parameters (43.79\wavenumber, 2.88\AA). This corresponds to the region where $ZPE<B$.]{\includegraphics[scale=0.48]{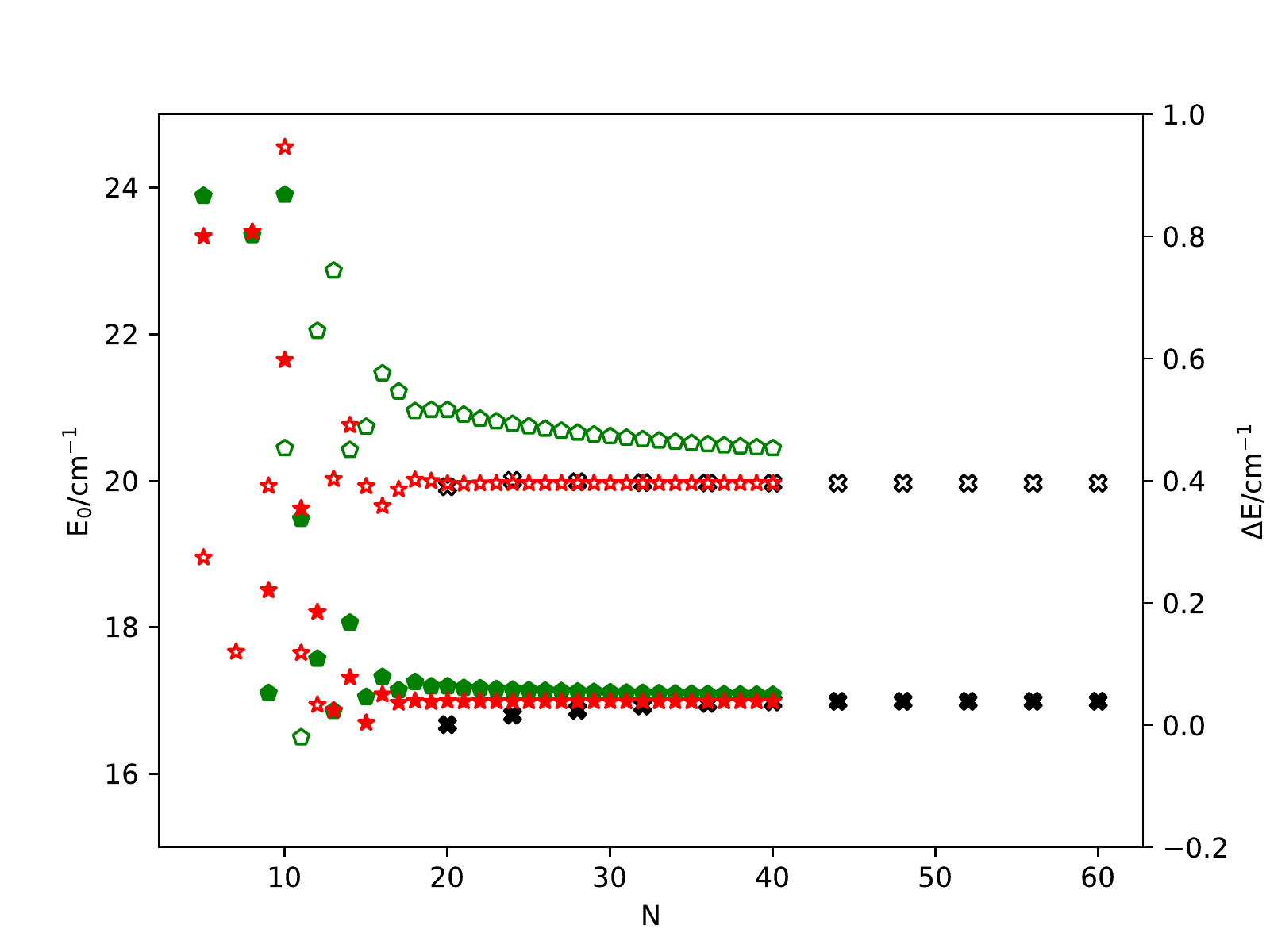}}
				\caption{	Convergence of the ground state energy and energy gap to the first excited state of a 1D Hamiltonian as in Equation \eqref{eq:LJ Hamiltonian} with $(\varepsilon, \sigma)$ Lennard-Jones parameters in (a) (43.79\wavenumber, 3.03\AA) and (b) (43.79\wavenumber, 2.88\AA). Symmetrised double minimum basis energies are shown as black crosses, HO DVR energies are red stars, and sinc DVR energies are green pentagons. Open shapes correspond to the ground state energy (left), and filled in shapes correspond to the energy gap to the first excited state (right).}
				\label{fig: DualMin vs Sinc Convergence}
			\end{figure}
		A one-dimensional potential was constructed by considering $V(0,0,z)$ for the potential given in Equation \eqref{eq:LJ Hamiltonian} for two different sets of LJ parameters, one which corresponds to the zero-point energy being similar to the barrier height and one where the barrier height is larger than the zero-point energy. The ground state energy is tracked against $N$ representing the number of quadrature points used. For both DVR techniques this is equivalent to the number of basis functions used, whereas the symmetrised double minimum basis set used $\frac{N}{2}$ functions. Both DVR basis sets were constructed by using a fixed interval of points, with $|z|<2$\AA.
		
		In both cases ($ZPE\approx B$ and $ZPE<B$), the sinc DVR basis set fails to converge to the ground state energy. In the former, it is much tighter and it appears that it will converge to the same energy but this is not as apparent in the latter. While both the HO DVR functions and symmetrised double minimum basis set converge the energy to sub milli-wavenumber accuracy, they achieve this in vastly different ways. When $ZPE\approx B$, the symmetrised double minimum basis set is flat, indicating that it had already converged in the smallest basis set size, but when $ZPE<B$, there is a small increase in energy moving from 24 points to 28 points, signifying the integrals had not fully converged. The HO DVR functions on the other hand (alongside the sinc DVR functions) have a fluctuating behaviour during the convergence. In the simpler polynomial oscillator, which lacks any energy barrier, one would expect the energy to decrease monotonically with increasing basis set size. Due to the presence of the double well, this is not the case and there is oscillatory convergence.
		
		As well as the ground state energy, another important quantity is the energy gap between the first excited state and ground state as it can be measured spectroscopically, corresponding to a tunnelling splitting. Considering that frequency of transitions is on the order\cite{mandziukQuantumThreeDimensional1994} of $10^1$\wavenumber, it is prudent to converge the energies and frequencies tightly. In both regions the sinc DVR basis frequency is larger than what is seen for the HO DVR functions and symmetrised double minimum basis set. The HO DVR transition frequency also suffers from oscillatory convergence in both regions, as it does for the ground state energy. The symmetrised double minimum basis set transition frequency also shows an initial rise as more quadrature points are needed to converge the higher energy state.
		
		We choose to use the symmetrised double minimum basis set for our calculations, as this takes into account features of the potential, does not require arbitrarily choosing the interval of sampling points and does not have oscillatory convergence of energy levels. We place 12 functions in each minimum for the $z$ direction, giving an overall basis size of 3456 functions. The Hamiltonian and overlap matrices were constructed using Equations \eqref{eq: Hmn} and \eqref{eq: Smn}. The basis set was then canonically orthogonalised\cite{lowdinNonorthogonalityProblemWork1970} by discarding linear combination of basis functions whose overlap eigenvalues were negligible.

	\subsection{Potential Energy Surface \label{sec:PES}}
		Of the three variable parameters in Equation \eqref{eq:LJ Hamiltonian}, $M$ was fixed to be the two-particle reduced mass of \Endo{Ne}{C70}. Given the form of the Hamiltonian, the quantity $M\varepsilon=\text{const}$ can be interpreted as an effective energy scale and therefore only one needs to vary. The LJ parameter space, $(\varepsilon,\sigma)$ chosen to explore was  $\varepsilon\in$ [20\wavenumber, 150\wavenumber] and $\sigma\in$ [2.85\AA, 3.05\AA]. These parameters were chosen as they ensured that the points for \Endo{Ne}{C70} given in Table \ref{table: Ne LJ parameters} were within it, and that the barrier height, $B$, along the $z$ direction varied over a substantial range: [1\wavenumber, 265\wavenumber]. This allows for regions where $\frac{ZPE}{B}<1$ and $\frac{ZPE}{B}>>1$. The location of minima also spanned a wide range: [0.35\AA, 0.85\AA] which leads to varying amounts of linear dependence within the basis set. 
		
		Allowing for a larger LJ parameter space can also introduce a double well in the $x$ and $y$ directions. These can be tackled with the same methodology outlined in Sections \ref{sec:Double Well} and \ref{sec:Comp}, but are not considered for the scope of this article. For each point in the LJ parameter space where the Hamiltonian was constructed and diagonalised, the PES was linearly shifted such that its minimum was set to zero.

			\begin{figure}
				\subfloat[Barrier height, $B$, in \wavenumber.]{\includegraphics[scale=0.48]{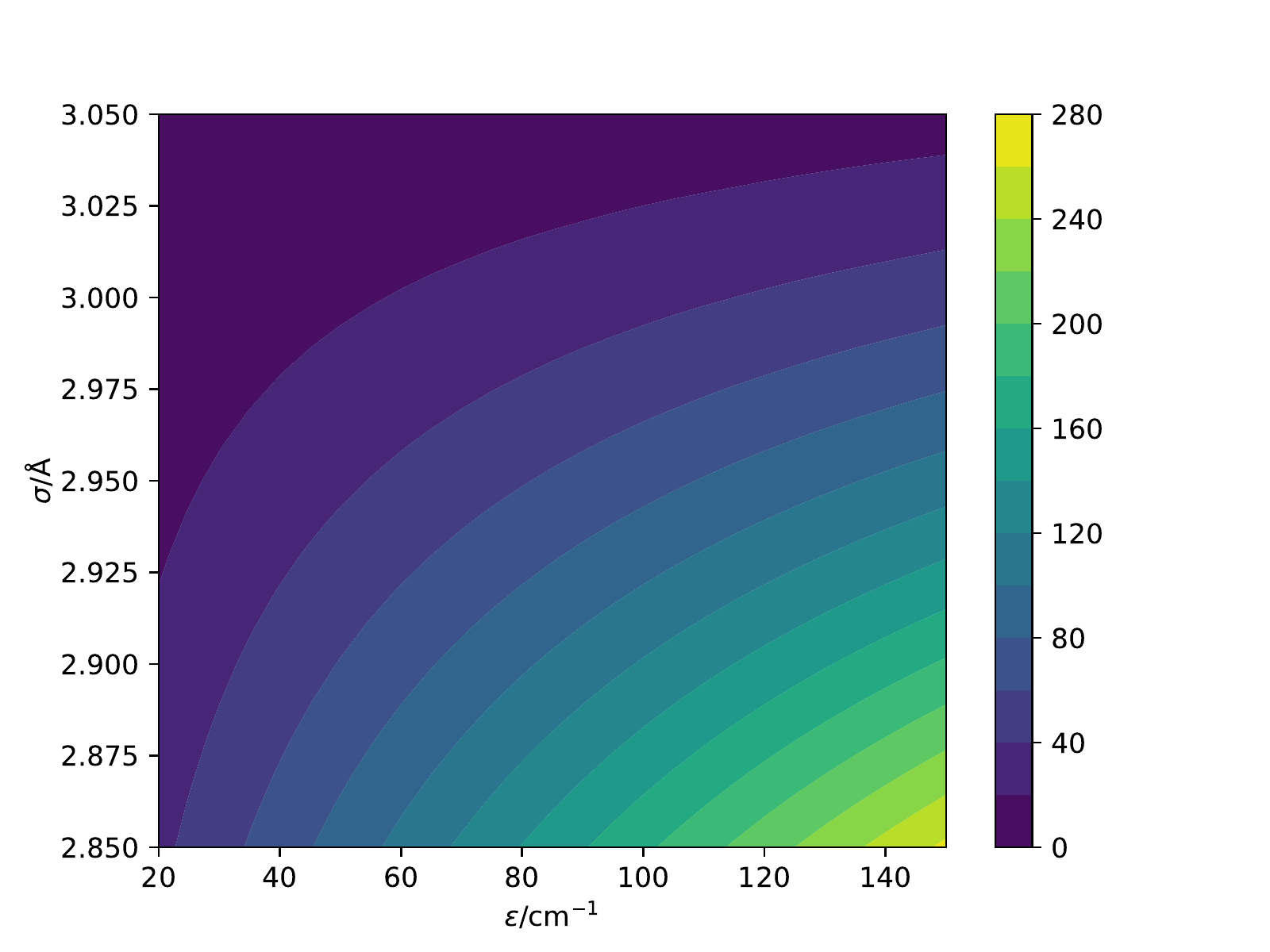} \label{fig: Barrier Height}}\\
				\subfloat[Location of $z_{\text{min}}$ in \AA]{\includegraphics[scale=0.48]{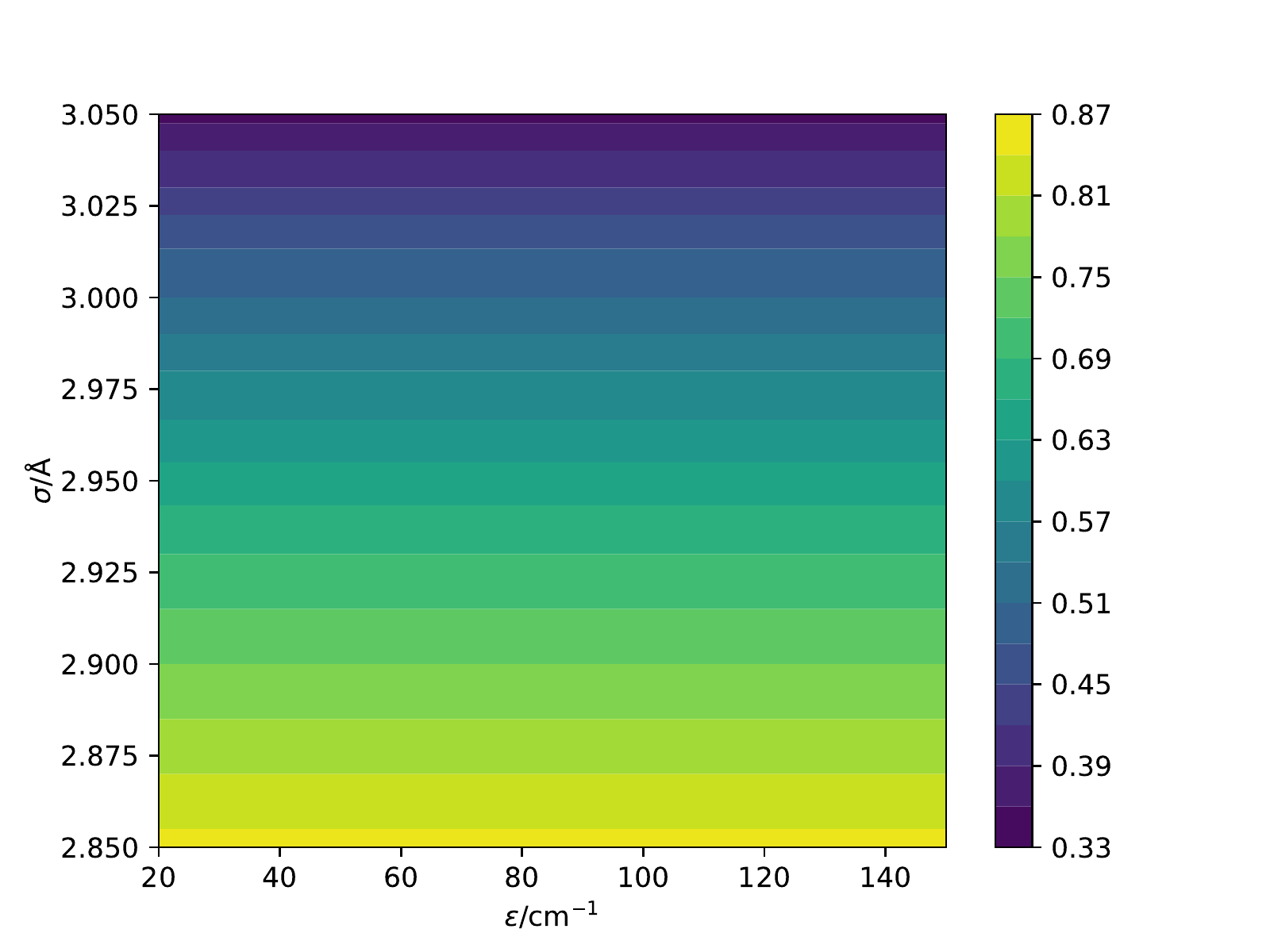} \label{fig: zmin Location}}
				\caption{Features of the double well, (a) barrier height and (b) position of minima, in the unique direction as defined in Figure \ref{fig: Fullertube Double Well} of the PES tracked across the LJ parameter space.}
				\label{fig: PES features}
			\end{figure}
		The location of the minima shows only $\sigma$ dependence, and as this increases, the minima slowly converge towards the origin. For the 1D LJ potential $r_{\mathrm{eq}}\propto \sigma$ and inside this fullerene cage this corresponds to pushing the endohedral species further away from the \ce{C} atoms and closer towards the centre. Once the minima merge together, increasing $\sigma$ further has no more effect as the centre of the cage is the furthest the species can be from the cage atoms.
		On the other hand, the barrier height shows dependence on both LJ parameters. While the relation with $\varepsilon$ is intuitive, increasing with larger $\varepsilon$ as raising the energy scale deepens the wells; increasing $\sigma$ reduces the barrier height. By pushing the minima closer together, they must eventually coalesce, eradicating the barrier completely. Due to this effective repulsive nature of increasing $\sigma$, once the minima merge the double well doesn't reappear.
			\begin{figure}
				\includegraphics[scale=0.48]{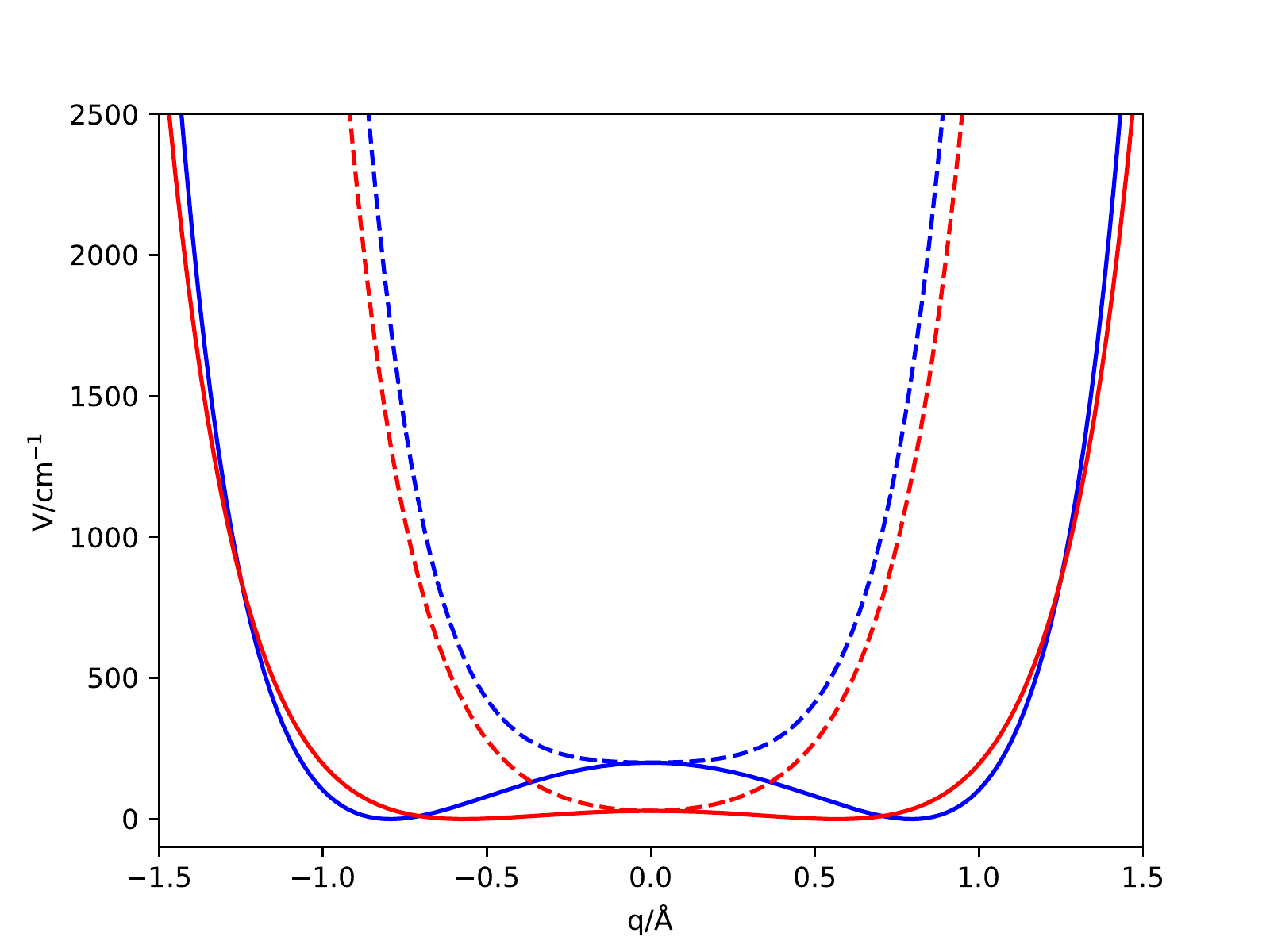}
				\caption{1D slices of Lennard-Jones potential along the $z$ and $x$ Cartesian directions. Solid line indicates $V(0,0,z)$ whereas dashed is $V(x,0,0)$. $V(0,y,0)$ is not shown as for this region it is almost identical to $V(x,0,0)$, differing only when the endohedral species is closer to the cage edge. The red curves have LJ parameters ($\varepsilon, \sigma)$ as (60, 2.98) and the blue curves have (140, 2.88).}
				\label{fig: 1D PES}
			\end{figure}

		The difference in 1D potentials throughout the LJ parameter space can be seen by considering two sets of parameters. In the $x$ (and $y$) directions, apart from the minimum having a different value, that of the barrier height, they both resemble potentials of an even polynomial oscillator as seen in Fig \ref{fig: 1D PES}. However, along the $z$ direction, not only are the minima found at different positions, namely 0.57\AA\, and 0.79\AA\, for the red and blue curves the barrier height also differs significantly changing from 30\wavenumber to 200\wavenumber respectively.
		
	\subsection{Ground State \label{sec:Ground State}}
		
		\subsubsection{Energy \label{sec:Evals}}
	
\begin{figure}
					\includegraphics[scale=0.48]{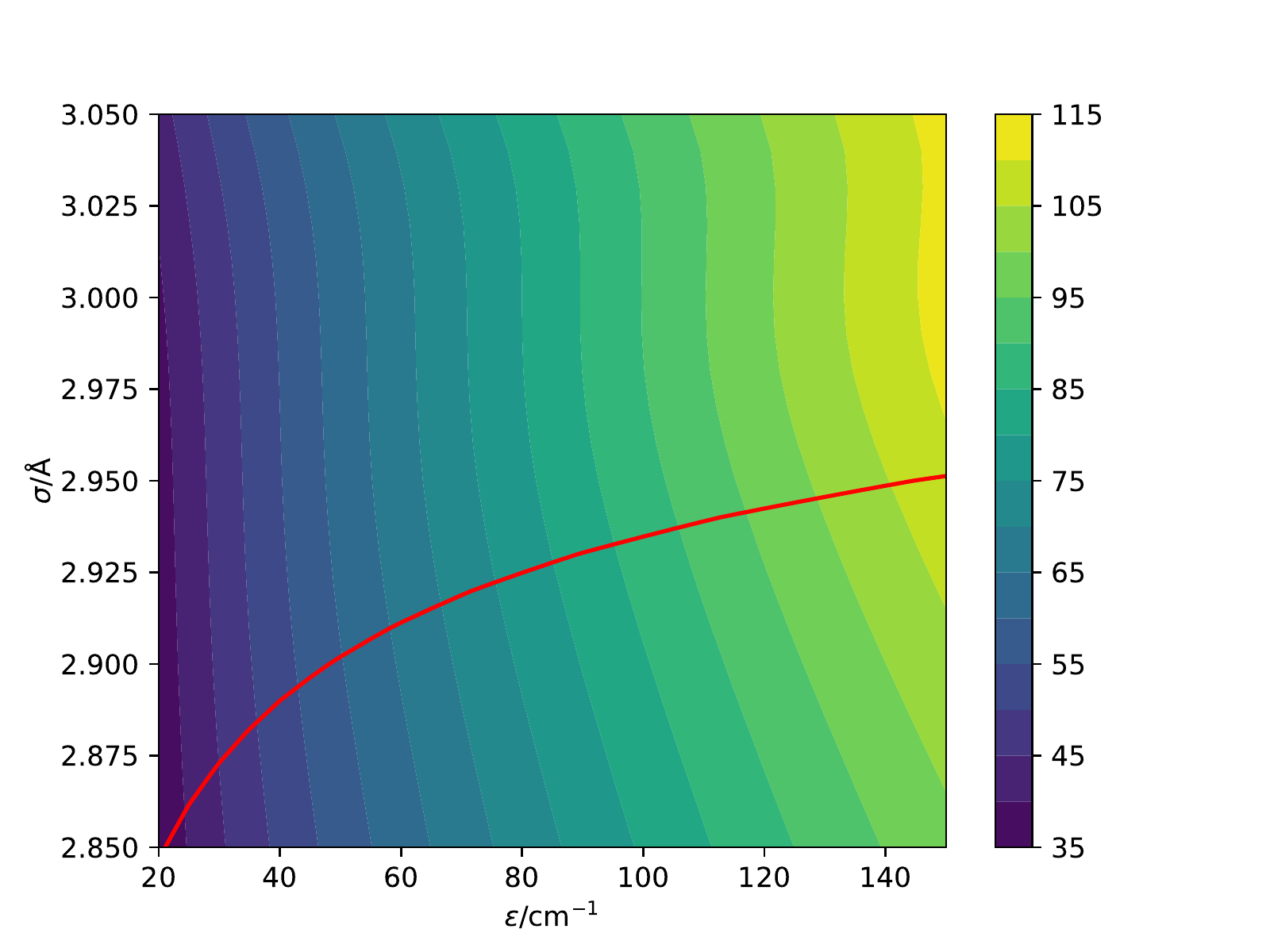}
					\caption{Ground state energy of \Endo{X}{C70} with Hamiltonian shown in \eqref{eq:LJ Hamiltonian} with $M$ corresponding to $\ce{^{20}Ne}$ in wavenumber across LJ parameter space. The red contour indicates the path where the barrier height $B$ and zero-point energy $ZPE$ are equal.}
					\label{fig: E0}
				\end{figure}

			The translational ground state of \Endo{X}{C70} was tracked across the aforementioned LJ parameter space and can be seen in Fig \ref{fig: E0}. The zero-point energy shows a strong dependence on $\varepsilon$, with the $ZPE$ getting larger as $\varepsilon$ increases. Increasing $\varepsilon$ corresponds to a more steeply growing potential as the endohedral atom moves further away from the origin. This increase in the effective force constant correlates with a larger frequency and $ZPE$ as $E\sim\omega\sim\sqrt{\frac{k_{\text{eff}}}{\mu}}$. The $ZPE$ also increases with increasing $\sigma$, but this is a much weaker dependence than seen for $\varepsilon$. Increasing $\sigma$ pushes the two minima closer together and lowers $B$, but this is compensated by the faster growing nature of the potential in the region $|z|>z_{\text{min}}$. As the size of this region increases, this acts as an increase in the effective force constant, pushing the $ZPE$ to be larger.
	
			The red contour indicates where $ZPE=B$ and partitions the LJ parameter space into two regions: above it, where $ZPE>B$, and below, where $B>ZPE$. Moving further away from this contour in this space, the more dominant $ZPE$ is above the contour and $B$ is below it. As $B$ dominates, the particle is more likely to be localised at the minima and the less likely it is to tunnel between them. As the $ZPE$ dominates, the more likely the particle is to be found in the centre as the double well is not a significant effect.

		\subsubsection{Moments \label{sec: Stats}}

\begin{figure}
					\subfloat[Standard deviation, $\sigma_z$ in \AA.]{\includegraphics[scale=0.48]{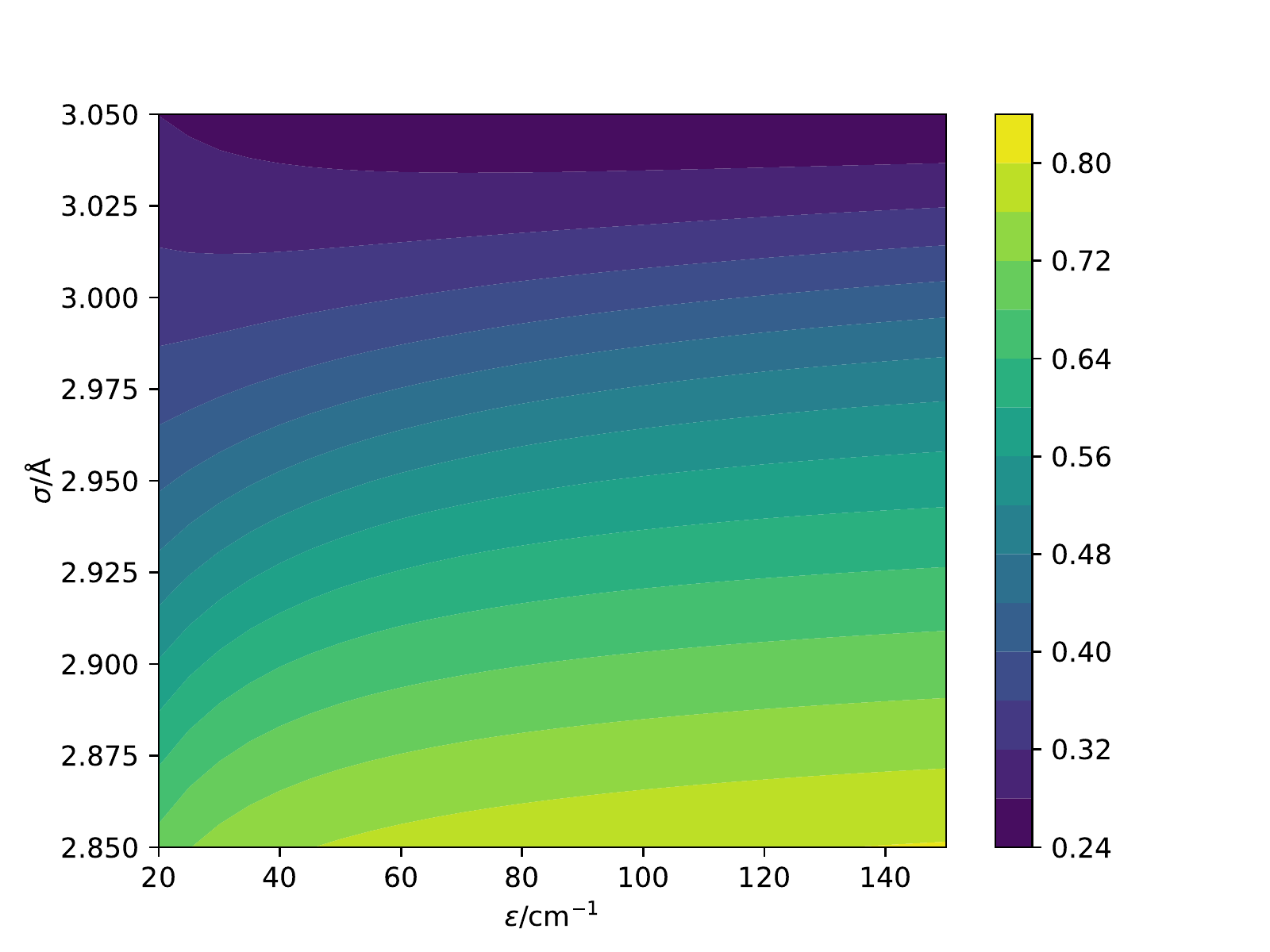}}\\
					\subfloat[Kurtosis, $\kappa_z$]{\includegraphics[scale=0.48]{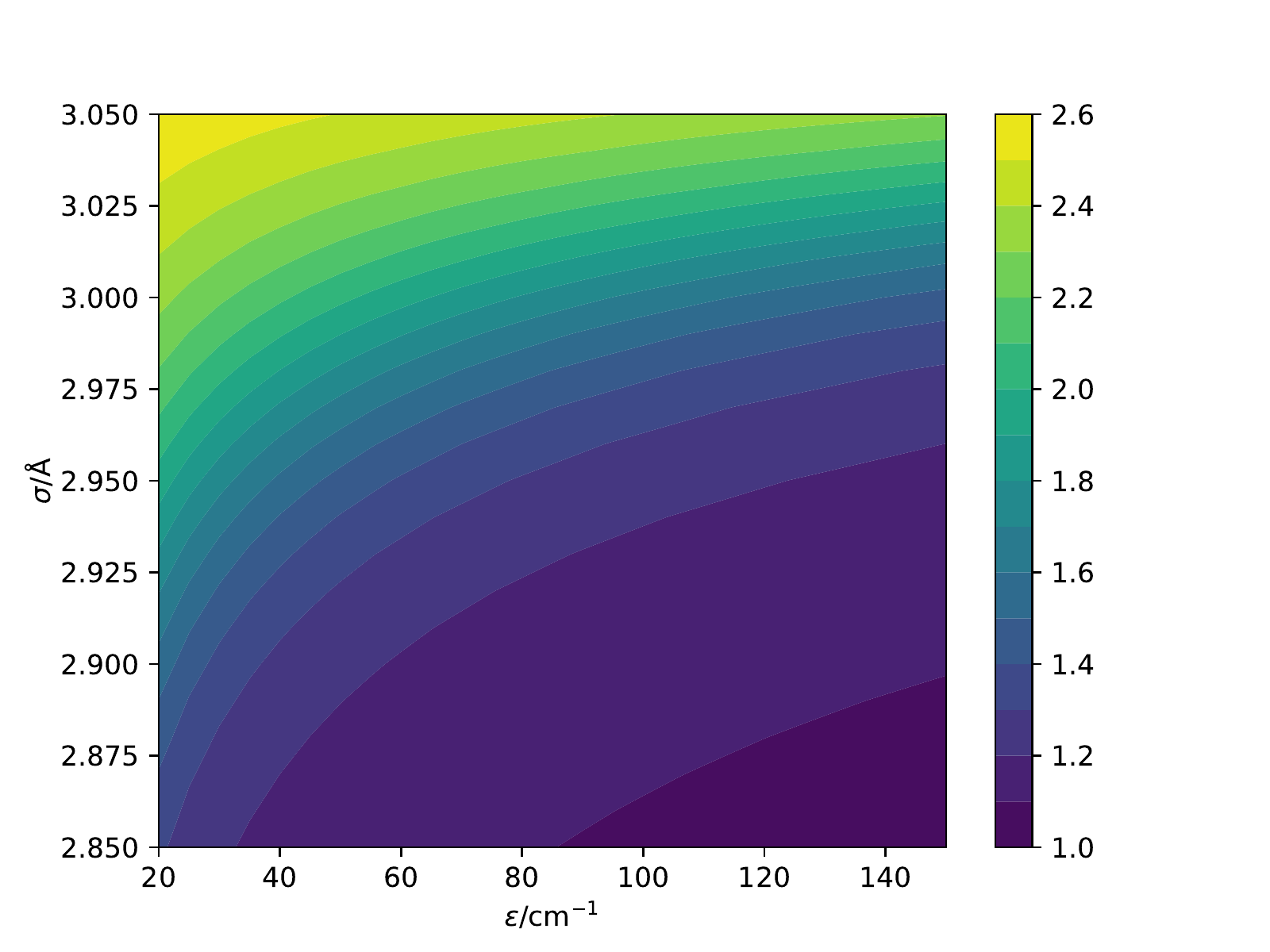}}\\
					\subfloat[Prolateness, $\varsigma$]{\includegraphics[scale=0.48]{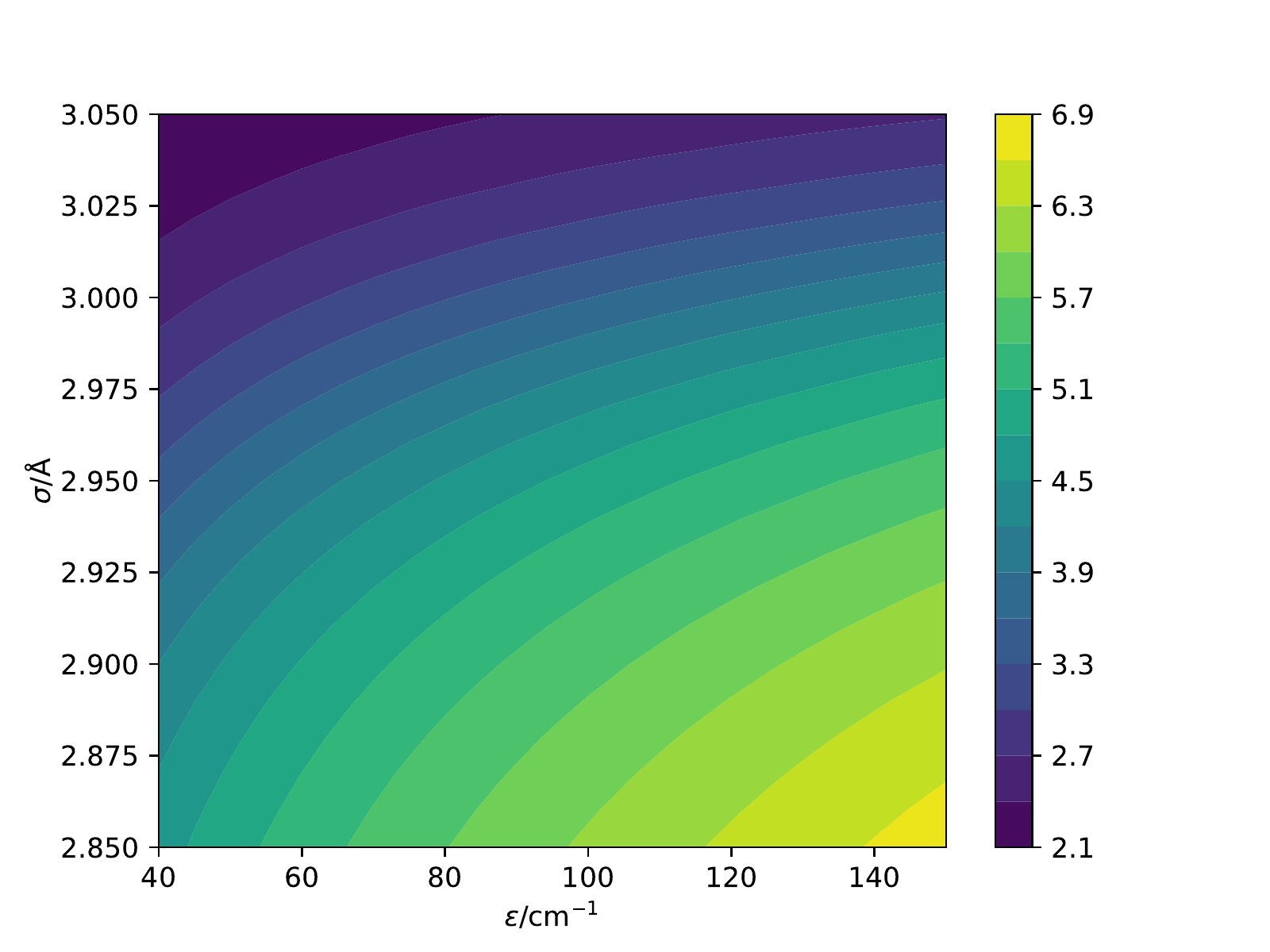}}
					\caption{Wavefunction moments, (a) standard deviation (b) kurtosis and (c) prolateness, of the ground state eigenfunction calculated along the unique $z$ axis across the LJ parameter space.}
					\label{fig: 2D Wavefunction Stats}
				\end{figure}

			Across the LJ parameter space, the standard deviation $\sigma_z$, the kurtosis $\kappa_z$, and standard deviation ratio $\varsigma$ as defined in Equations \eqref{eq: Variance Ratio} and \eqref{eq: Kurtosis} in Section \ref{sec:Wavefunc Stats} were calculated. The standard deviation and kurtosis were tracked throughout the parameter space and shown in Fig \ref{fig: 2D Wavefunction Stats}. The standard deviation lies in the interval [0.24\AA, 0.81\AA], and this value is always smaller  than the position of the minima seen in Fig \ref{fig: zmin Location}. Acknowledging that $\braket{z}=0$ by symmetry this indicates that the endohedral atom is more likely to be located between the two minima than closer to the cage. While the overall skewness of the wavefunction must also be zero due to symmetry, when the wavefunction exhibits two maxima each of those is individually skewed towards the centre of the cage. This is due to the potential experiencing a local maximum at the origin, seen in Fig \ref{fig: 1D PES}, instead of growing steeply towards the cage.
			 
			The standard deviation increases with decreasing $\sigma$ and with increasing $\varepsilon$. Both of these effects change the nature of the minima in the potential, by either pulling them apart and further away from the centre or by accentuating their depth. The influence of the minima is to drag wavefunction density towards themselves and away from the origin which leads to an increase in the standard deviation.

			As mentioned in Section \ref{sec:Wavefunc Stats}, the standard deviation cannot distinguish between a wavefunction that exhibits a single maximum at the origin or two maxima near the minima in the potential. This is determined by the kurtosis, defined in \eqref{eq: Kurtosis}. For this system, the kurtosis lies in the interval [1.07, 2.58] implying all these wavefunctions represent a platykurtic distribution, one where the kurtosis is lower than a Gaussian. The kurtosis decreases with increasing $\varepsilon$ and decreasing $\sigma$. Varying the LJ parameters in this way pulls the minima apart and emphasises their depth. This drags more wavefunction density away from the origin, in the region $|z|>\sigma_z$ which puts more density in the tails and this implies a decrease in kurtosis.

			To fully classify the 3D wavefunction, instead of just considering $\sigma_z$, the quantity $\varsigma$ can be used instead as this is the ratio of standard deviations in the $z$ and $x$ directions which gives a measure of the prolateness of the ellipsoidal wavefunction, as defined in Equation \eqref{eq: Variance Ratio}.  This quantity increases with increasing $\varepsilon$, as the weight of the minima in the potential increases, pulling density towards themselves, and it decreases with increasing $\sigma$, as the minima get closer together. 	The prolateness is considered alongside $\kappa_z$, defined in Equation \eqref{eq: Kurtosis}, which gives an indication of how much the idealised single-peaked Gaussian wavefunction has been squashed, pushing density further away from the origin.

				\begin{figure}
					\subfloat[Contours in LJ parameter space used to track the ground state wavefunction statistics defined in Equations \eqref{eq: Variance Ratio} and \eqref{eq: Kurtosis}.]{\includegraphics[scale=0.48]{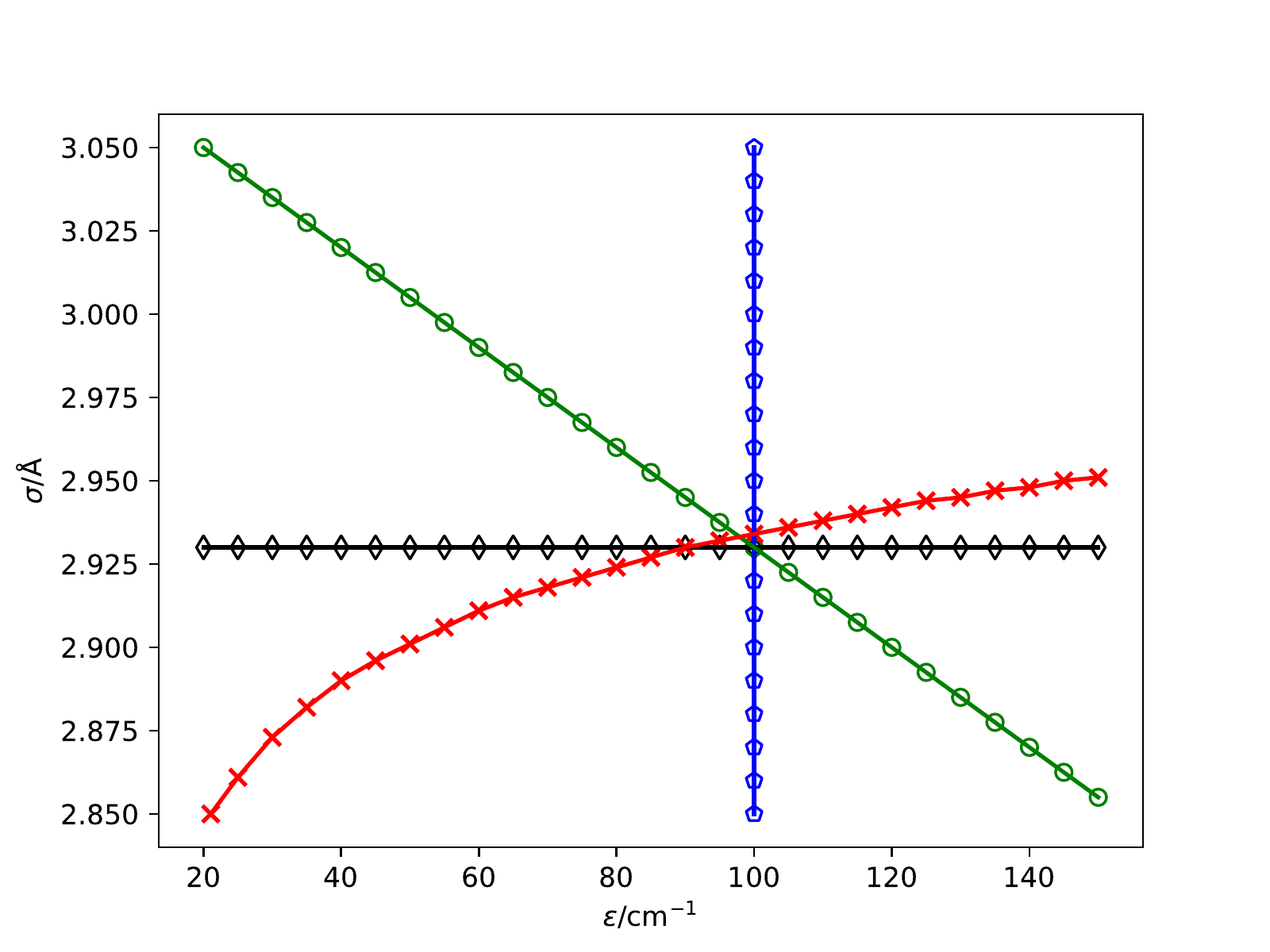}\label{fig: LJ paths}}\\
					\subfloat[Wavefunction statistics across 4 contours in LJ parameter space.]{\includegraphics[scale=0.48]{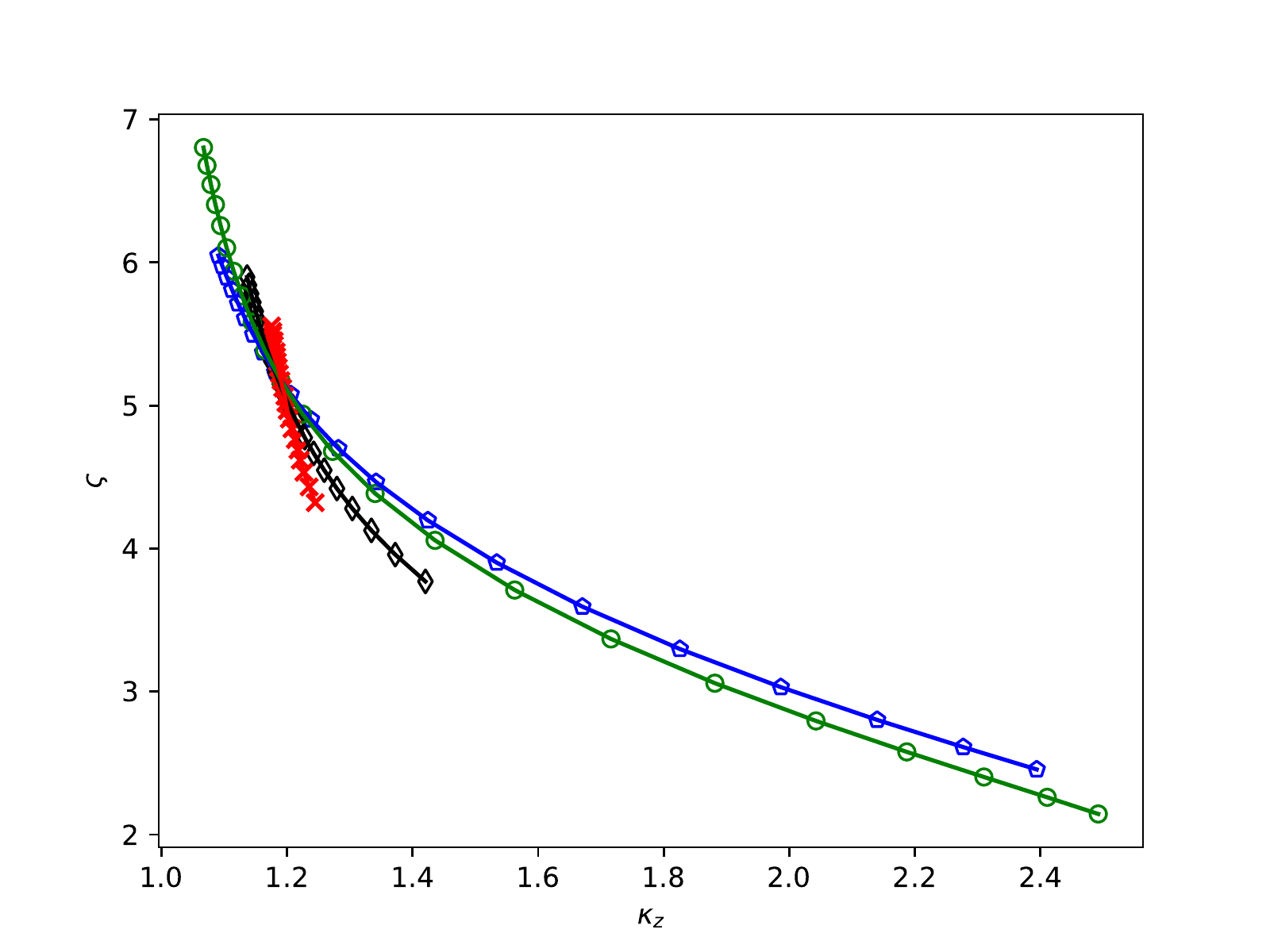} \label{fig:Ne stats}}
					\caption{Ground state wavefunction prolateness and kurtosis in (b) tracked along the contours in the Lennard-Jones parameter space in (a). The black line is at fixed $\sigma = 2.93$\AA, the blue line is at fixed $\varepsilon=100$\wavenumber\, the green line is along the path $\sigma=3.08-0.0015\varepsilon$ and the red curve is along the $ZPE=B$ contour shown in Fig \ref{fig: E0}.}
					\label{fig: LJPaths and NeStats}
				\end{figure}

			Four paths were chosen to exhibit these statistics on the ground state wavefunction for the \Endo{X}{C70} system. The first two were for fixed $\varepsilon$ and fixed $\sigma$, which isolate the effect of a single LJ parameter on the shape of the wavefunction. A third path is one where both LJ parameters vary, along the line $\sigma=3.08-0.0015\varepsilon$ in order to check whether the effects of a single LJ parameter compound or cancel each other. The last path is along the $ZPE=B$ contour, as the range of these two statistics can determine which of them is a good measure of double-peakedness.
			
			Keeping either $\varepsilon$ or $\sigma$ fixed and varying the other LJ parameter, $\varsigma$ lies in the interval [2.45,6.05]. With increasing $\sigma$ and decreasing $\varepsilon$, the prolateness $\varsigma$ decreases indicating a move to a more spherical wavefunction. This effect is compounded when both quantities are varied as $\varsigma$ spans a larger range. Along $B=ZPE$, the dependence of $\varsigma$ on both LJ parameters remains the same but the range is tightened.

			The kurtosis on the other hand has the opposite dependence on the LJ parameters than the standard deviation ratio. The range spanned by varying $\varepsilon$ is smaller than when varying $\sigma$ as changing the latter has a stronger effect on whether the two maxima in the wavefunction eventually coalesce. This is the same behaviour as the two minima in the potential; as they get closer together the wavefunction moves towards a single peak. Once again both these effects stack up when both parameters are varied as $\kappa_z$ approaches its largest value of 2.58 which is only slightly smaller than for a pure Gaussian. Along the $ZPE=B$ contour, as for $\varsigma$, the range for $\kappa_z$ is very narrow.

			The reason to use $\kappa_z$ as a more informative measure on the type of wavefunction becomes apparent when looking along the $ZPE=B$ contour where $\varsigma$ is in the interval [3.76,5.56] but $\kappa_z$ is in the interval [1.17,1.31]. Only when both LJ parameters are varied does the kurtosis reach values above 2.2 illustrating the move to a single peaked wavefunction. These statistics vary smoothly as the LJ parameters are varied, indicating the wavefunction shape also smoothly changes throughout the parameter space. The quantity $\varsigma$ changes to show how stretched the ellipsoidal shape is along the unique axis, whereas $\kappa_z$ illustrates the deformation of the idealised Gaussian, single maximum wavefunction into two equivalent peaks. This occurs by effectively squashing the single peak of a standard Gaussian and stretching it outwards such that the maxima occur near the potential minima. This changeover occurs around a kurtosis of 2.2. As these statistics completely define the wavefunction shape, this can then be used to re-construct the PES and the LJ parameters of the \ce{X-C} interaction can be discerned.

		\subsubsection{Nuclear Orbitals \label{sec:Orbitals}}

				\begin{figure*}
					\subfloat[(40\wavenumber, 2.890\AA)]{\includegraphics[scale=0.48]{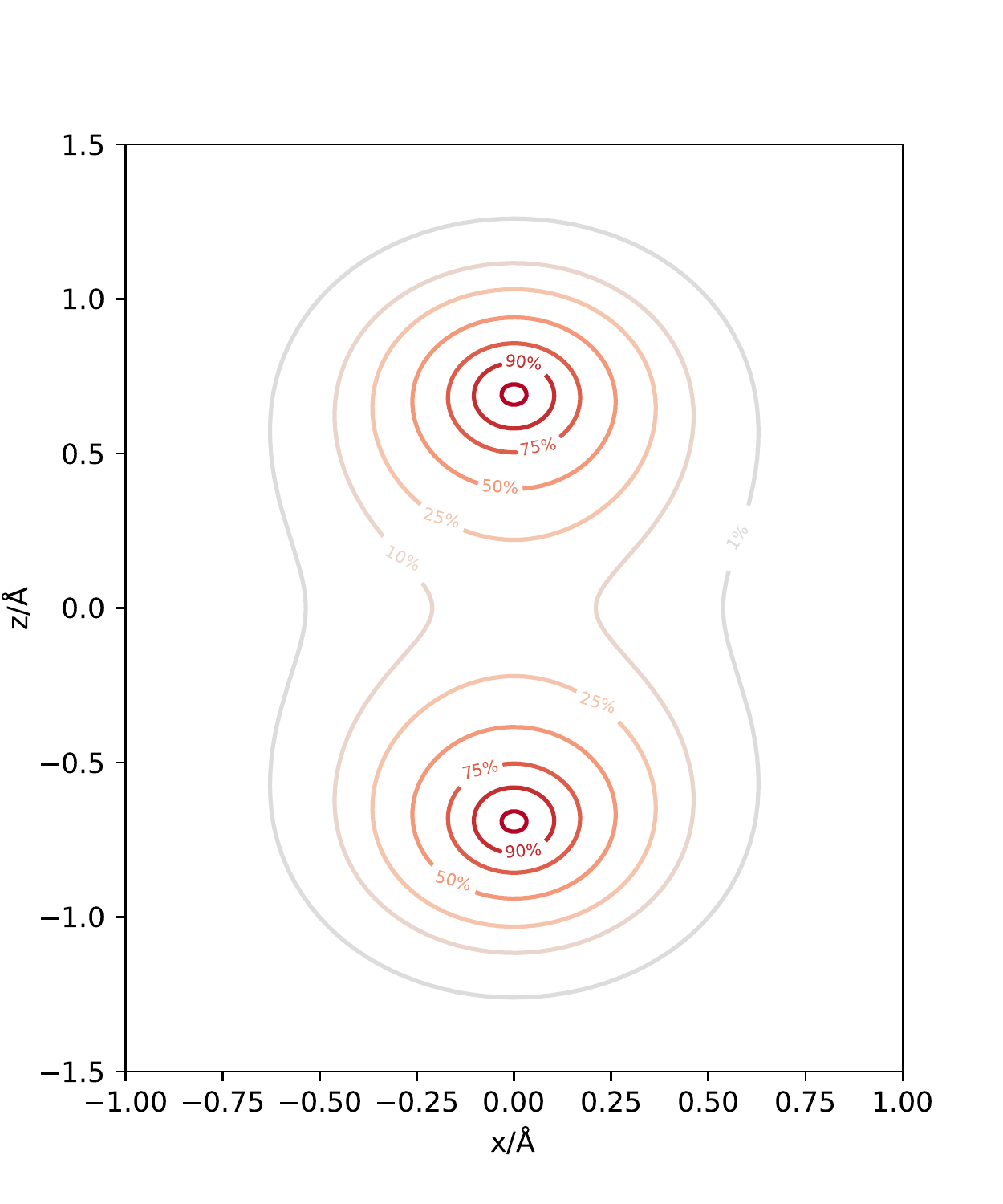}}\,
					\subfloat[(70\wavenumber, 2.918\AA)]{\includegraphics[scale=0.48]{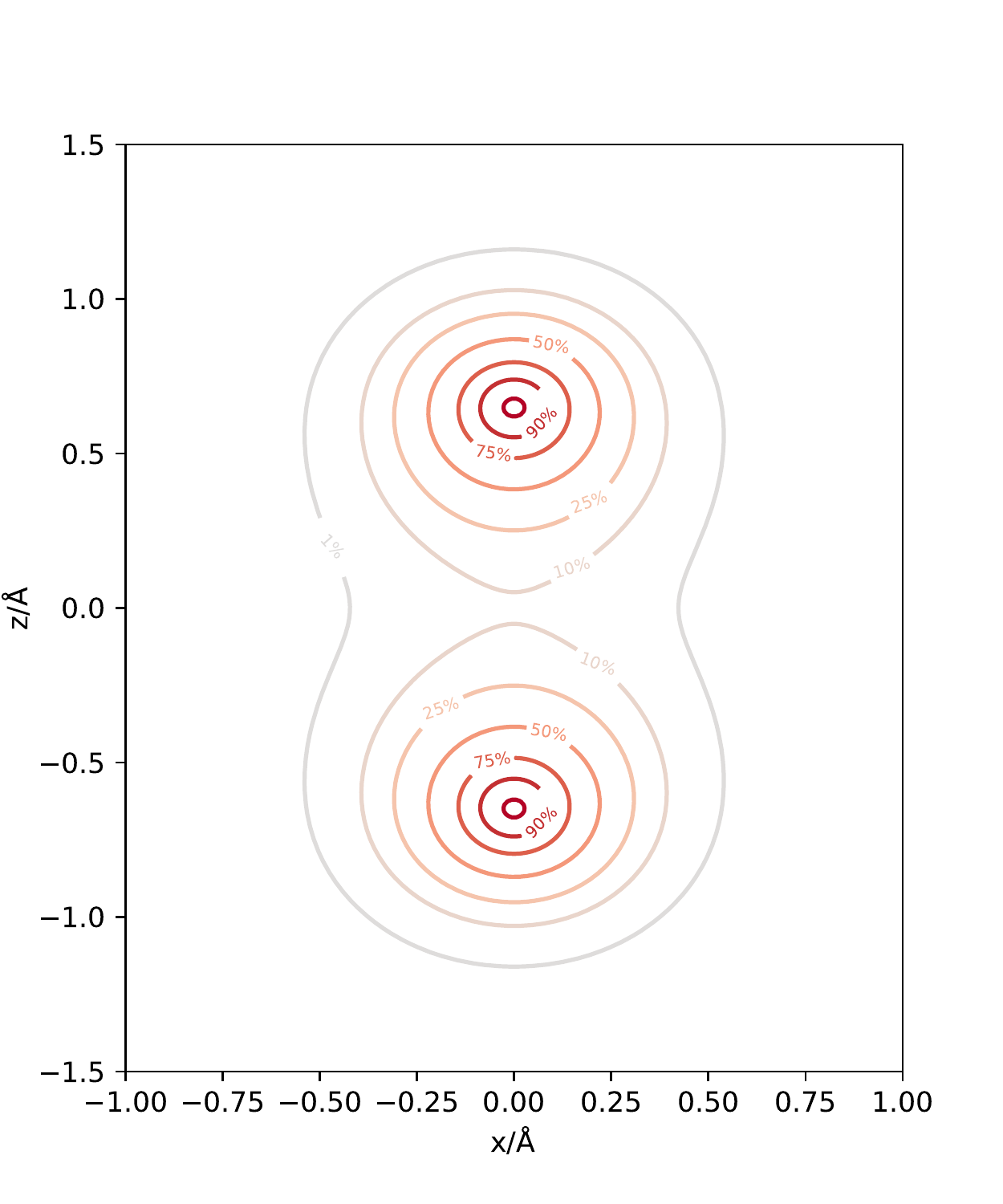}}\\
					\subfloat[(110\wavenumber, 2.938\AA)]{\includegraphics[scale=0.48]{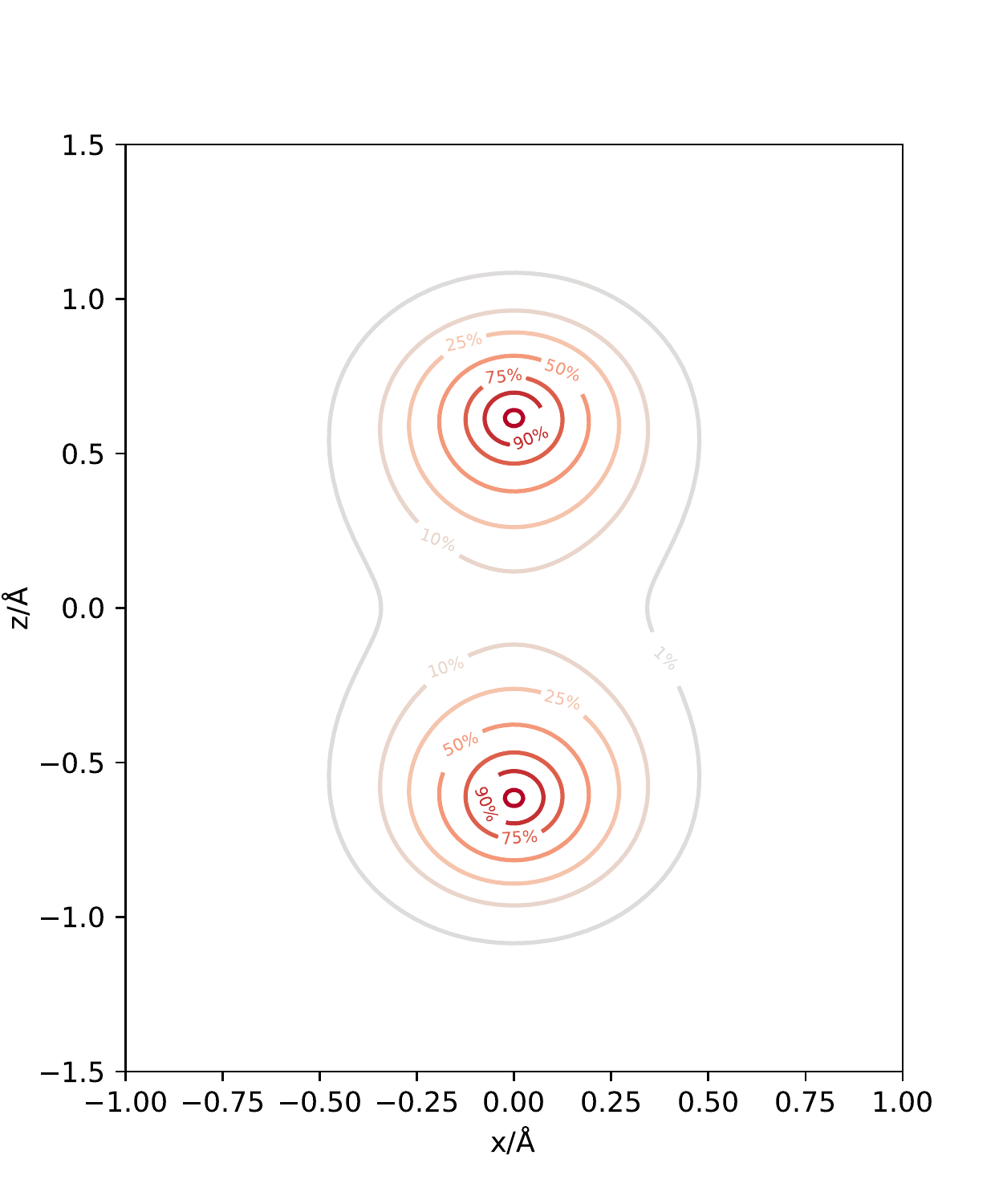}}\,
					\subfloat[(150\wavenumber, 2.951\AA)]{\includegraphics[scale=0.48]{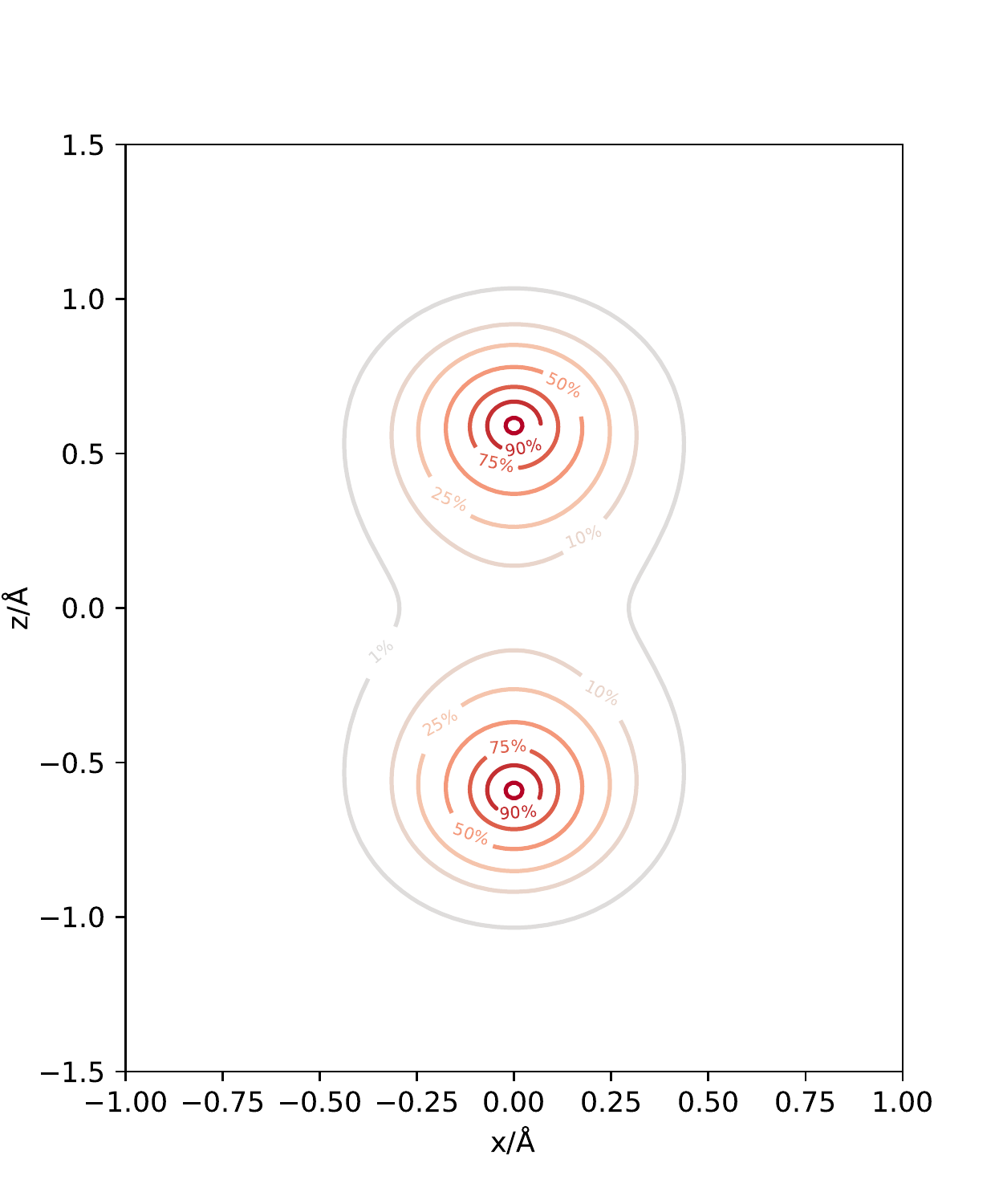}}
					\caption{$xz$ contours of the ground state wavefunction with $y=0$ along the $B=ZPE$ path for the $(\varepsilon, \sigma)$ parameters (a) (40\wavenumber, 2.890\AA), (b) (70\wavenumber, 2.918\AA), (c) (110\wavenumber, 2.938\AA), and (d) (150\wavenumber, 2.951\AA). The contours are taken at 1\%, 10\%, 25\%, 50\%, 75\%, 90\%, and 99\% of the maximum amplitude of the wavefunction.}
					\label{fig: Ne E0B Wavefunc}
				\end{figure*}

			In order to give more tangible meaning to the wavefunction statistics described in Section \ref{sec: Stats}, the wavefunctions themselves can be visualised. This gives more of an intuition as to what exactly $\varsigma$ and $\kappa_z$ correspond to. Along the $ZPE=B$ path, there is not too much discernible change in the wavefunction contours seen in Fig \ref{fig: Ne E0B Wavefunc}. As $\varepsilon$ and $\sigma$ increase, the wavefunction becomes more contracted. This is weakly due to $\sigma$ as this pushes the minima closer to the centre, from 0.78\AA\, down to 0.66\AA. The stronger dependence is on  $\varepsilon$, as this dictates the depth of the well and steepness of the potential, and the larger it is the more tightly bound the wavefunction will be.

\begin{figure*}
					\subfloat[(40\wavenumber, 3.02\AA)]{\includegraphics[scale=0.48]{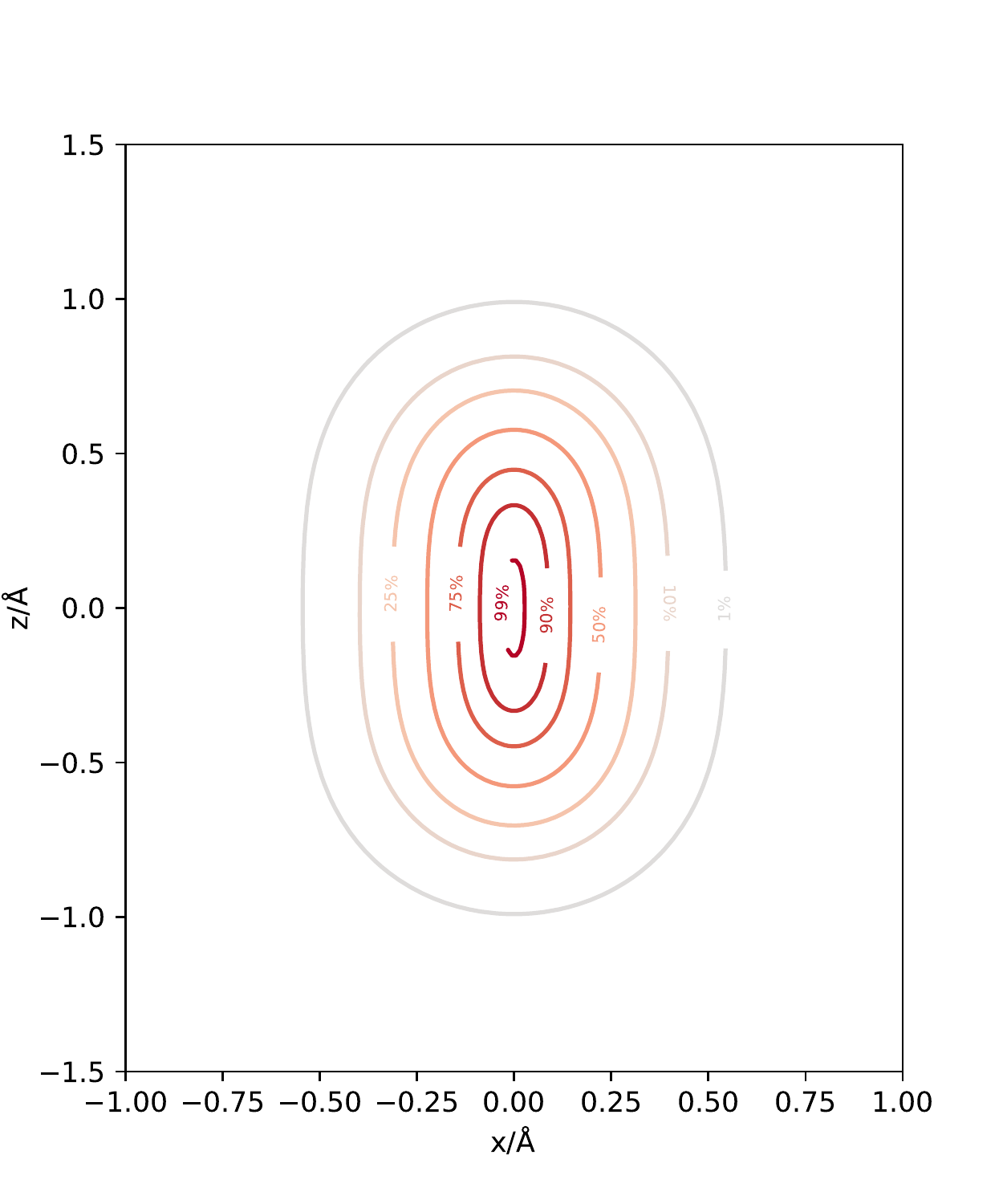}}\,
					\subfloat[(70\wavenumber, 2.975\AA)]{\includegraphics[scale=0.48]{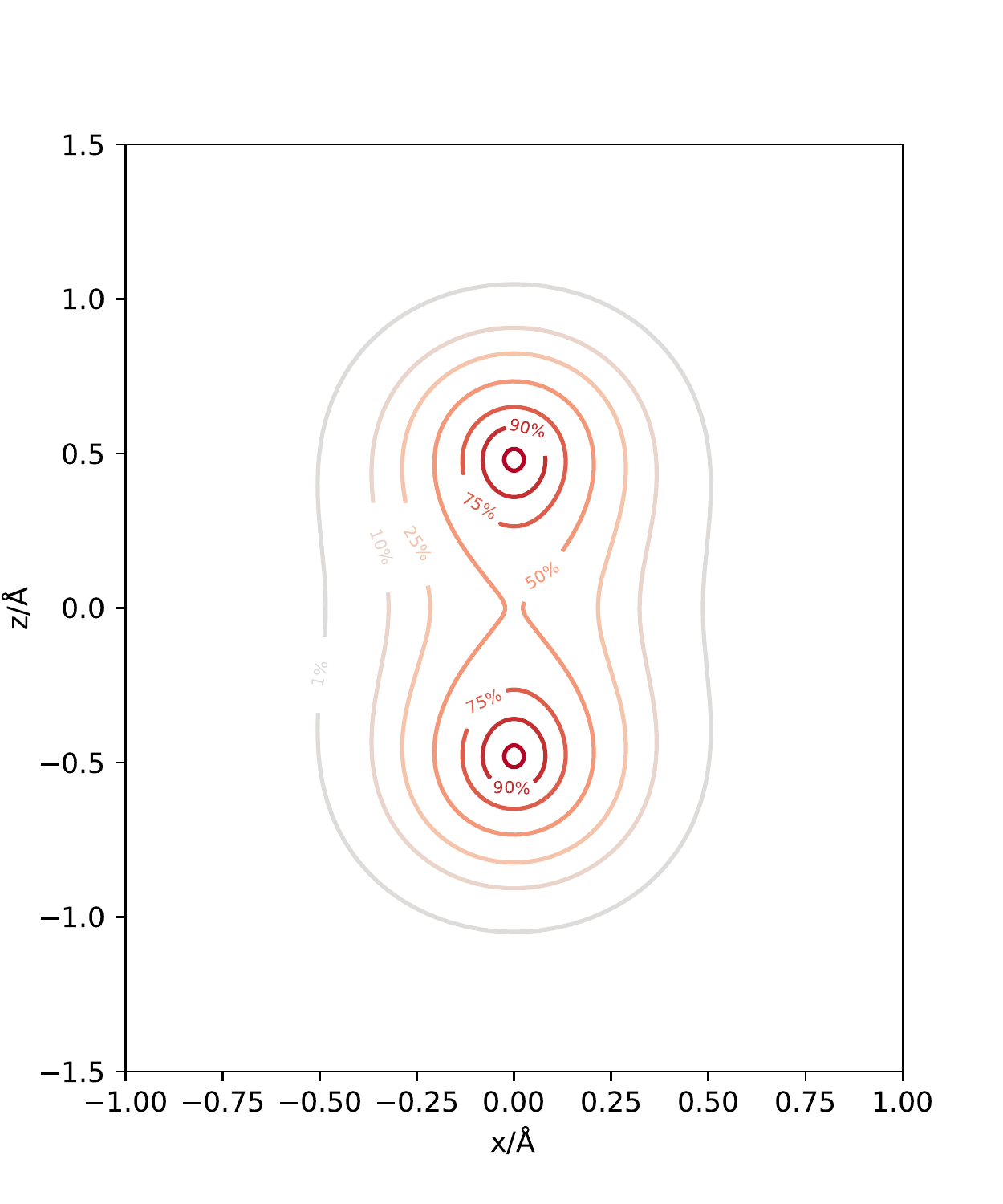}}\\
					\subfloat[(110\wavenumber, 2.915\AA)]{\includegraphics[scale=0.48]{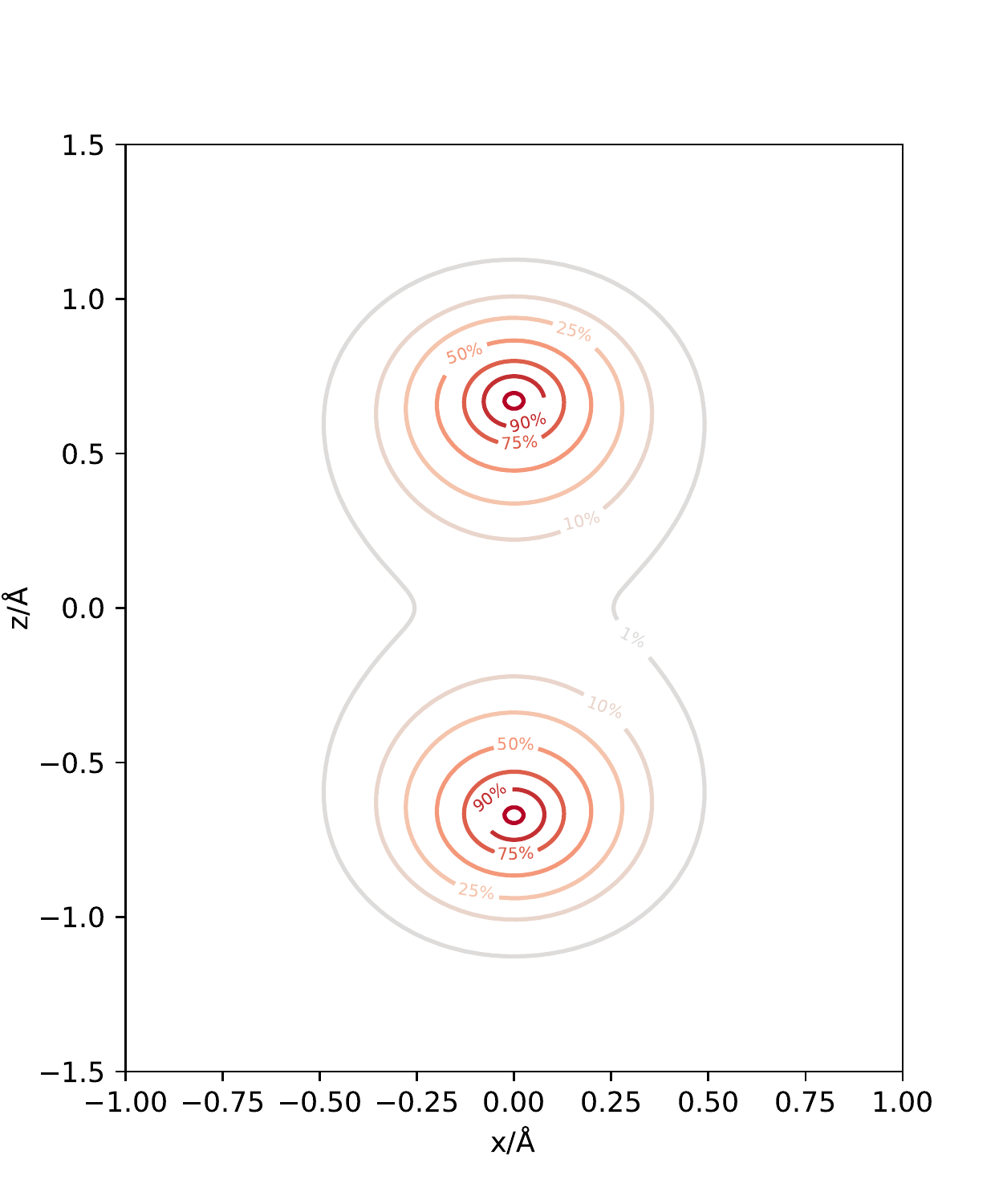}}\,
					\subfloat[(150\wavenumber, 2.855\AA)]{\includegraphics[scale=0.48]{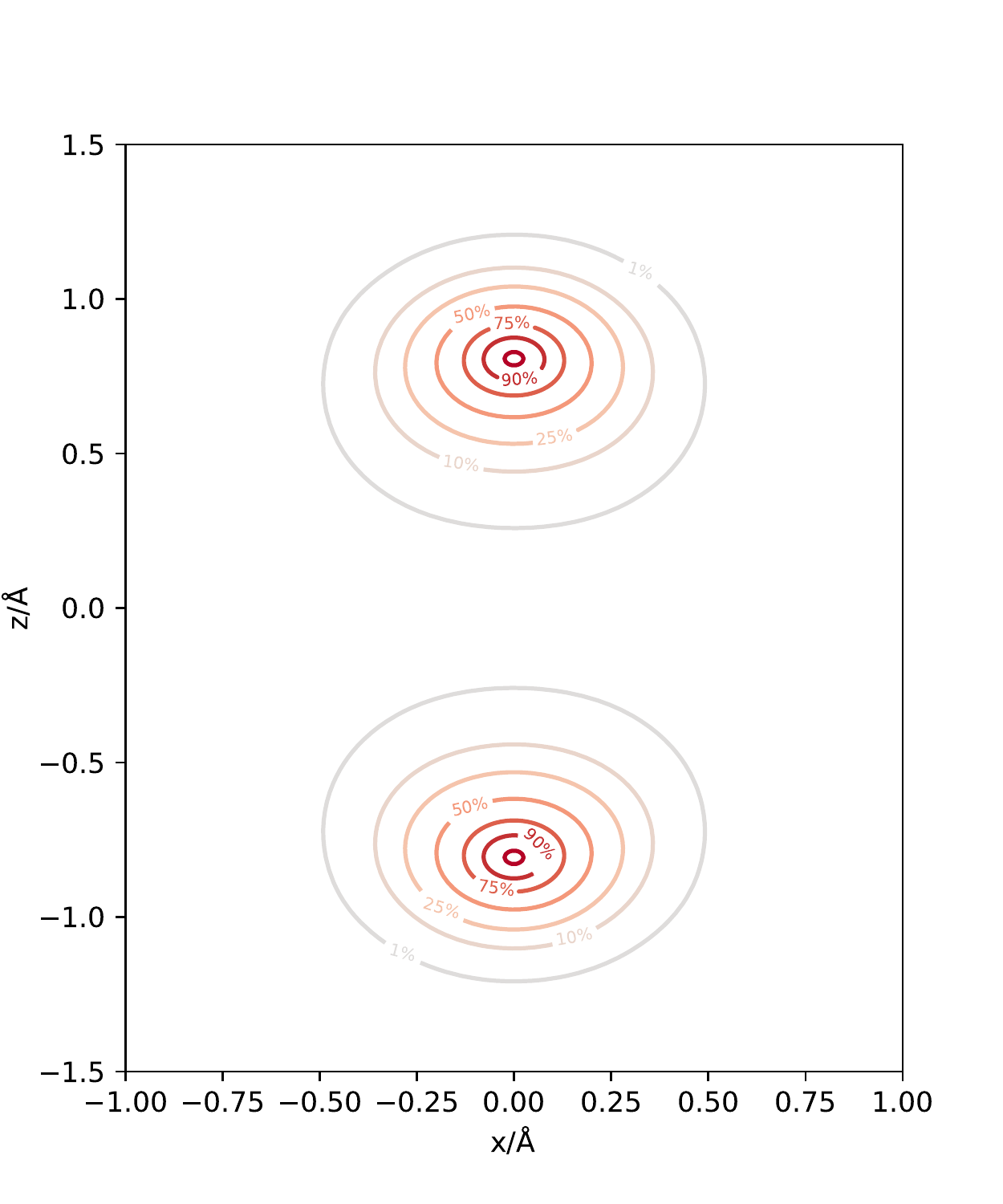}}
					\caption{$xz$ contours of ground state wavefunction with $y=0$ along the path $\sigma=3.08-0.0015\varepsilon$ for the $(\varepsilon, \sigma)$ parameters (a) (40\wavenumber, 3.02\AA), (b) (70\wavenumber, 2.975\AA), (c) (110\wavenumber, 2.915\AA), and (d) (150\wavenumber, 2.855\AA). The contours are taken at 1\%, 10\%, 25\%, 50\%, 75\%, 90\%, and 99\% of the maximum amplitude of the wavefunction.}
					\label{fig: Ne vary LJ Wavefunc}
				\end{figure*}

\begin{figure}
				\includegraphics[scale=0.48]{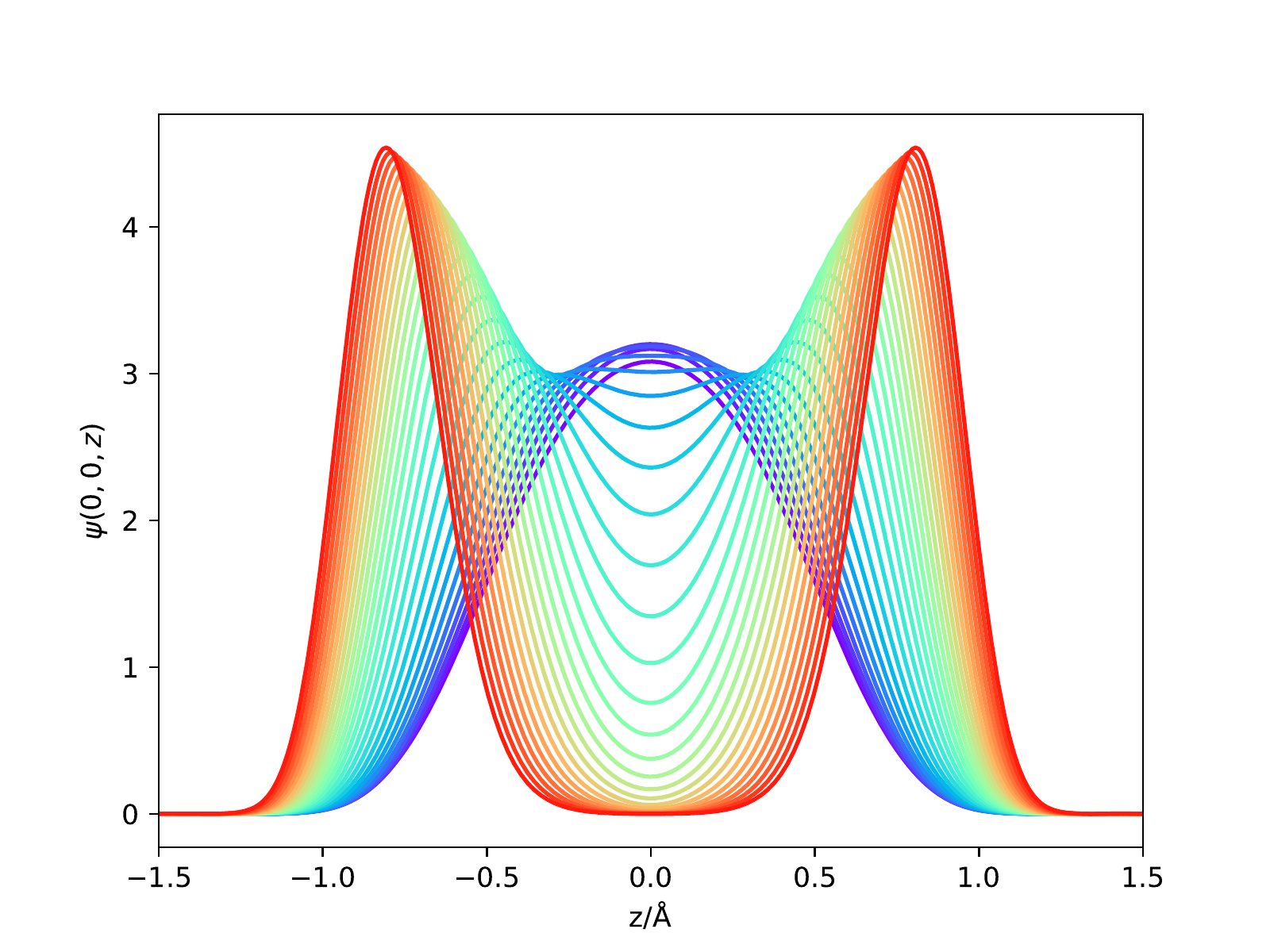}
				\caption{$\psi(0,0,z)$, on LJ path $\sigma=3.08-0.0015\varepsilon$ with $\varepsilon$=20\wavenumber in purple, to 150\wavenumber in red.}
				\label{fig: 1Dz Wavefunc}
			\end{figure}

		The stark difference in wavefunction shapes can be seen along the line which varies both LJ parameters, as illustrated in Figs \ref{fig: Ne vary LJ Wavefunc} and \ref{fig: 1Dz Wavefunc}. As $\varepsilon$ increases, and $\sigma$ decreases on the line $\sigma=3.08-0.0015\varepsilon$, there is a smooth deformation of the 1D wavefunction from a more typical Gaussian, which becomes squashed and stretched along the line into two shoulder peaks and eventually into two very distinct, localised maxima. Considering the 2D contour plots, when $(\varepsilon, \sigma$) are (40, 3.02), the wavefunction exhibits a typical, simple ellipsoidal character.  Moving to (70, 2.975), while maintaining a mostly overall ellipsoidal shape, there are clearly two maxima present in this wavefunction. When the LJ parameters are (110, 2.915), the two maxima are very prominent and the wavefunction shows a significant peanut-like shape. Moving to (150, 2.855), this is the extreme case showing two isolated lobes of wavefunction density, indicating two prominent maxima in the minima of the potential, with no discernible wavefunction amplitude connecting these regions.

		Both $\varsigma$ and $\kappa_z$ span a wide interval along for these wavefunctions. Between the first two wavefunctions $\varsigma$ is fairly similar, only increasing slightly but the kurtosis drops quickly from 2.5 down to 2.0. For the latter two, the kurtosis is nearly the same, but $\varsigma$ increases from 5.5 to 6.8. The decrease in kurtosis and increase in standard deviation ratio both correspond to moving wavefunction density away from the origin. The latter is more sensitive initially on the move from a single peak wavefunction to having two maxima. Once these maxima are prominent, putting more density there has less significant effect on the kurtosis as this is lower bounded by 1, but $\sigma_z$ keeps on increasing.

	\subsection{Excited States \label{sec:Excited States}}
	
		In addition to converging the ground state, excited state energies were also calculated. Here, instead of varying the LJ parameters which were fixed to be (43.79\wavenumber, 3.03\AA), the parameters for \Endo{Ne}{C70} as shown in Table \ref{table: Ne LJ parameters}, the fullerene cage was varied along the $D_{5d/h}$ series from \ce{C70} to \ce{C100}. The lowest 50 eigenstate energies were calculated and assigned quantum numbers $(L,m,n_z)$, analogously to Refs \citenum{xuCoupledTranslationrotationEigenstates2009} and \citenum{mandziukQuantumThreeDimensional1994}, where $L$ and $m$ are the angular momentum quantum numbers of the 2D isotropic harmonic oscillator in the $xy$ plane and $n_z$ the quantum number of the oscillator in the $z$ direction. 
		
		\begin{figure}
			\includegraphics[scale=0.48]{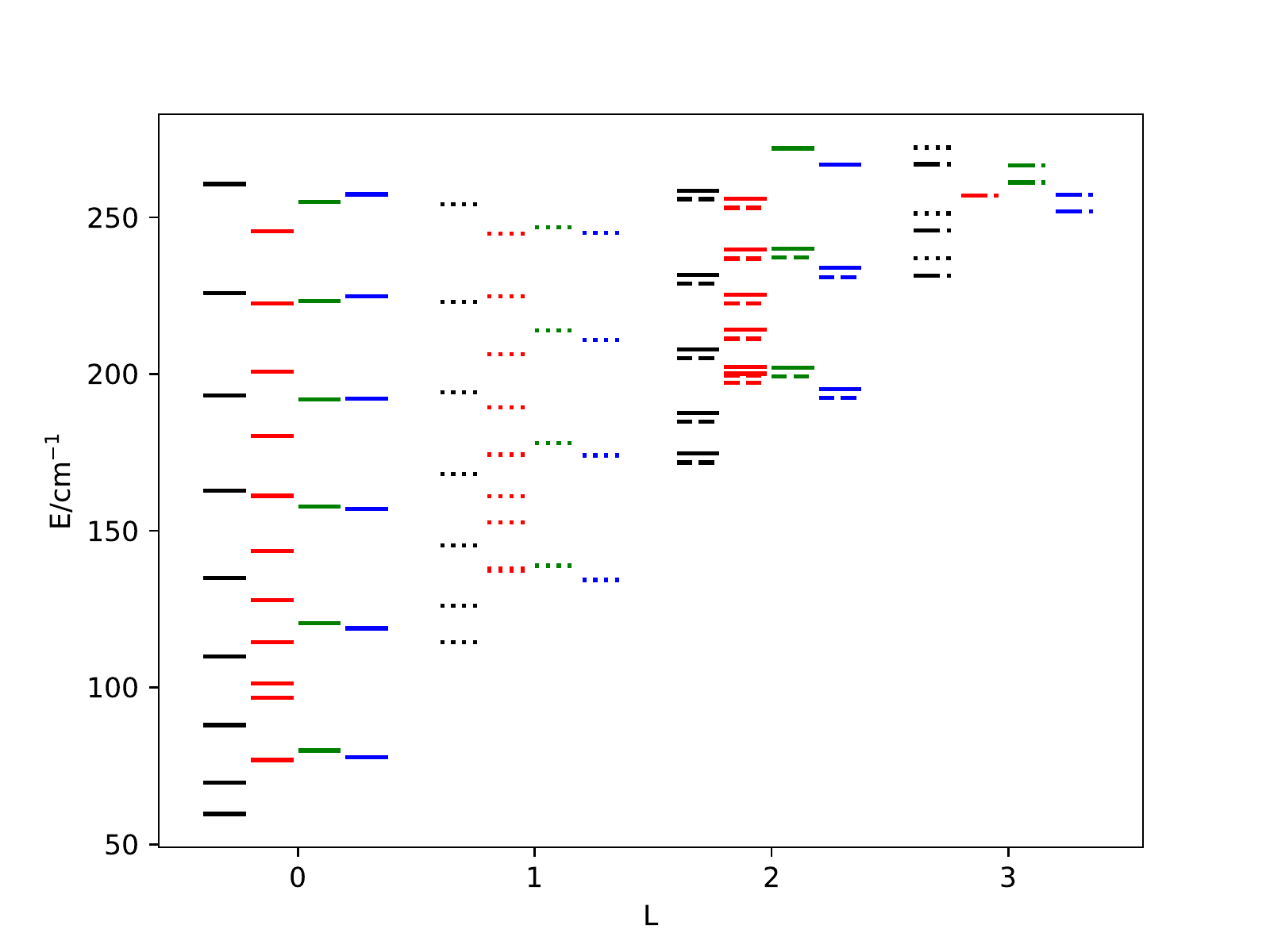}
			\caption{Energy levels of lowest 50 eigenstates of \ce{X}@\ce{C}$_{\mathrm{n}}$ in \wavenumber, with fixed LJ parameters ($\varepsilon, \sigma)$ as (43.79\wavenumber, 3.03\AA). \ce{C70} is in black, \ce{C80} in red, \ce{C90} in blue, and \ce{C100} in green. States are indexed with quantum numbers $(L,m,n_z)$. $m=0$ shown with solid lines, $|m|=1$ as dotted lines, $|m|=2$ as dashed lines, and $|m|=3$ as dot-dashed lines. States in \ce{C90} and \ce{C100} appear as near-degenerate pairs with quantum numbers $(L,m,n_z)$ and $(L,m,n_z)+1$ for $n_z$ even.}
			\label{fig: Fullertube Quantum Numbers}
		\end{figure}
		
		These quantum numbers are assigned by analysing the nodal structure of the wavefunctions, with $n_z$ being the number of nodes along the $z$ direction. By considering the $\frac{L-m}{2}$ radial and $m$ angular nodes, the angular momentum quantum numbers $L$ and $m$ can also be assigned. In Figure \ref{fig: Fullertube Quantum Numbers}, the eigenstates of \Endo{X}{C70} are in black, \Endo{X}{C80} in red, \Endo{X}{C90} in green and \Endo{X}{C100} in blue. States with $|m|>0$ are doubly degenerate with quantum numbers $\pm m$. States with $m=0$ are solid lines, $|m|=1$ are dotted lines, $|m|=2$ are dashed lines and $|m|=3$ are dot-dashed lines. 
		
		For fixed $(L,m)$, going up in energy corresponds to increasing $n_z$. For a fixed $L$ and $n_z$, states increase in energy with decreasing $|m|$, with the restriction that $|m|\leq L$. Once the fullertube reaches \ce{C90}, the minima are far enough apart that for all these eigenstates the wavefunctions are not interacting and all the states appear as $(L,m,n_z)$ and $(L,m,n_z+1)$ degenerate pairs. 
		
		The lifting of degeneracy of states with differing $|m|$ for a fixed $L$ indicates the anharmonic nature of the potential. Across the set of fullerenes, this anharmonicity is negative, as seen in \Endo{H2}{C70}\cite{xuCoupledTranslationrotationEigenstates2009} due to the higher degree nature of the nearly radially symmetric potential. Along the $z$ direction, while for \Endo{X}{C70} this also shows negative anharmonicity, the longer fullerenes show a slightly different pattern. Because of the larger barrier height, these states are more influenced by the double well, and so in \Endo{X}{C80}, the negative anharmonicity only starts from the $(0,0,2)$ state, with an unusually large energy difference between the $(0,0,1)$ and $(0,0,2)$ states seen in Fig \ref{fig: Fullertube Quantum Numbers}, which could be rationalised as the endohedral atom almost ``breaking-free'' of the double well influence.
		
		For \Endo{X}{C90} and \Endo{X}{C100}, after accounting for the $(0,0,n_z)$ and $(0,0,n_z+1)$ states arising in degenerate pairs for $n_z$ even, the states seem to show a small positive anharmonicity, below an energy of 250\wavenumber. This is because as the states lie below the barrier height, the potential almost mimics what would be seen for a diatomic stretch. In that limit, the diatomic dissociates forcing positive anharmonicity on the potential. But in this case, this is the case only until the atom is in a high enough energy state as to not feel the effect of this double well, and then the negative anharmonicity returns.
		
\begin{figure}
			\subfloat[$(0,0,0)\rightarrow(0,0,1)$ transition energy in \wavenumber. The red contour indicates where the barrier height and zero point energy are equal and the yellow contour indicates a transition energy of 1\wavenumber.]{\includegraphics[scale=0.48]{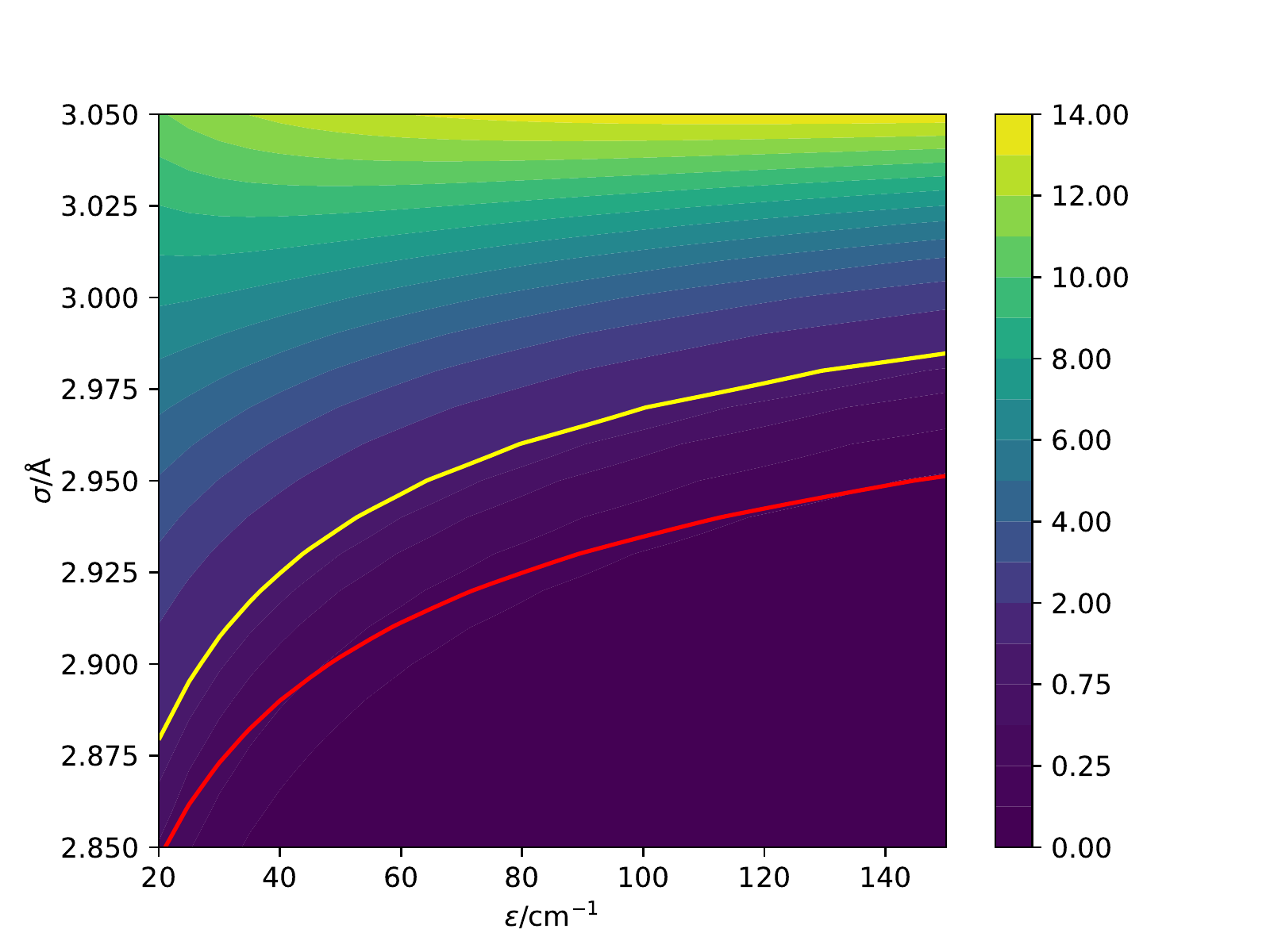}}\\
			\subfloat[$(0,0,0)\rightarrow(1,1,0)$ transition energy in \wavenumber]{\includegraphics[scale=0.48]{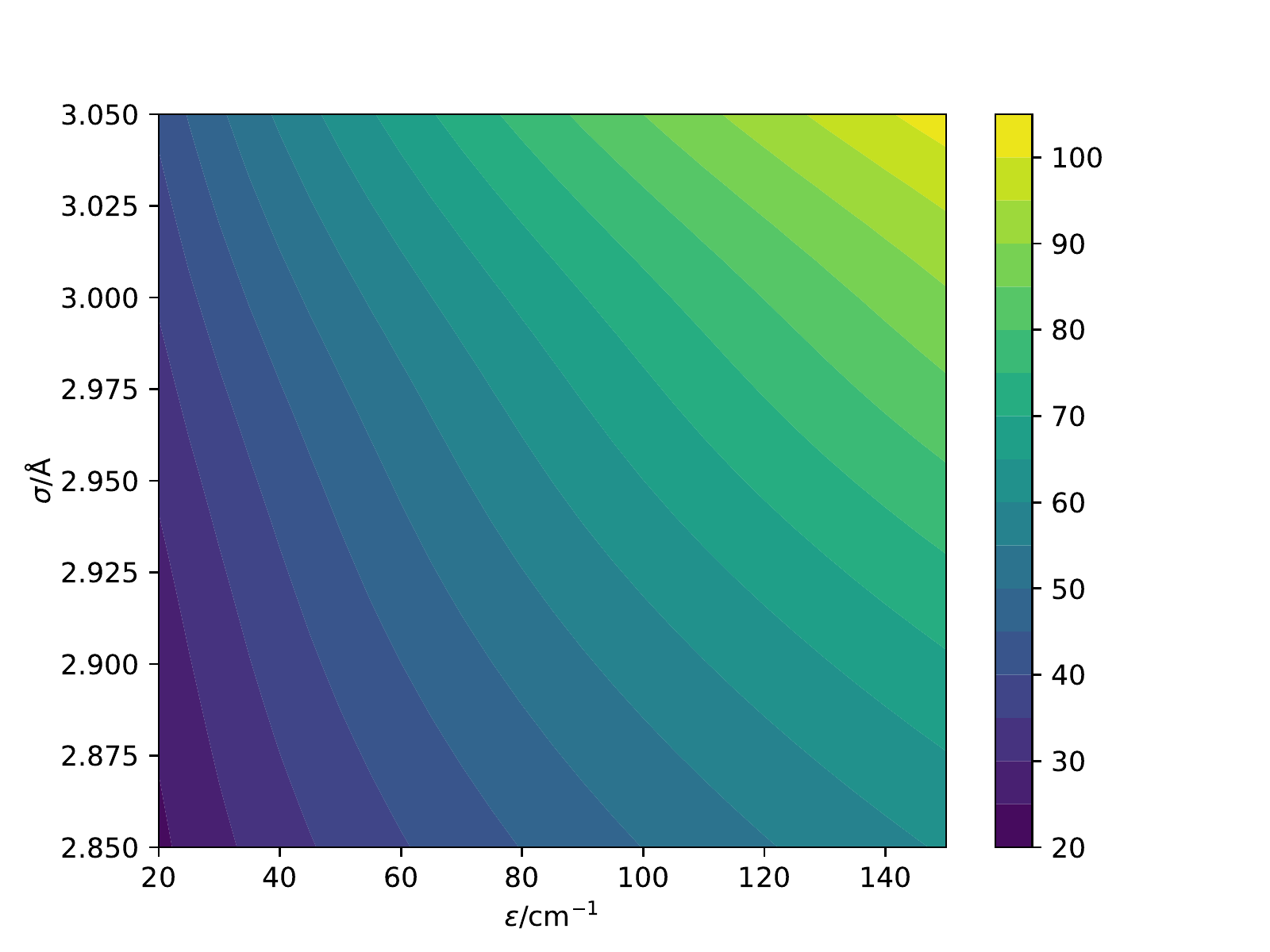}}
			\caption{Transition energies of the fundamental transition in the (a) anistropic $z$ direction and (b) in the 2D isotropic $xy$ plane.}
			\label{fig: Fundamental Transitions}
		\end{figure}

		Traversing the $D_{5d/h}$ series, the transition energies between corresponding states decreases. This can be rationalised by considering Figure \ref{fig: Fullertube Double Well}. As the fullerene is elongated, the double wells become deeper hence the particle is more attracted in the local minima. Spectroscopically, the transition energies between states can be measured. If this is conducted at a low enough temperature, only the ground state will be sufficiently populated for transitions to be visible, and so the fundamental transitions $(0,0,0)\rightarrow(0,0,1)$ and $(0,0,0)\rightarrow(1,1,0)$ as seen in Figure \ref{fig: Fundamental Transitions} are of importance. These correspond to the fundamental frequency of the 1D oscillator in the $z$ direction and 2D isotropic oscillator in the $xy$ plane for \Endo{X}{C70} respectively.
		
		The $xy$ fundamental transition shows a stronger dependence on $\varepsilon$, as this is directly related to the ``stiffness'' of the isotropic oscillator and a larger stiffness implies a larger frequency. The dependence on $\sigma$ is less intuitive, but as this is the length scale of the \ce{X-C} interaction, the larger the value, the ``stronger'' the interaction leading to a larger frequency. As can be see in Figure \ref{fig: Fullertube Quantum Numbers}, the number of $(0,0,n_z)$ states lying below this first degenerate state differs depending on the position of the minima and barrier height which can also suggest a larger frequency.
		
		The fundamental transition, which in this case is always the transition in the $z$ direction, shows very different characteristics to the $xy$ fundamental. The dependence on $\varepsilon$ is very weak as this modulates the depth of the double wells, seen in Figure \ref{fig: Barrier Height}. There is a stronger dependence on $\sigma$, which characterises the distance between the minima as seen in Figure \ref{fig: zmin Location} and as $\sigma$ increases, the minima get closer together increasing the interaction of the wavefunctions in both wells leading to a larger tunnelling splitting. The red contour indicates where the barrier height is equal to the zero point energy, whereas the yellow contour shows a frequency of 1\wavenumber, which is near the resolution of what could be seen experimentally. Spectroscopically assigning the transitions, the LJ parameters for the \ce{X-C} interaction can be determined.
		
		To compare the varying nature of these fundamental transitions across the LJ parameter space to the $D_{5d/h}$ fullertube series, a quantity to consider would be the ratio of the fundamentals. However, due to the double well and small tunnelling splitting, the ratio of $xy$ fundamental and $z$ fundamental is a not a well behaved quantity as the $z$ fundamental approaches 0. Instead, the $z$ fundamental could be replaced by the $(0,0,1)\rightarrow(0,0,2)$ transition, the $z$ hot band. In Fig {fig: Fundamental Transitions}, this is the transition what looks as if its the fundamental, and is better behaved. For the cases where $ZPE>B$, as for \Endo{X}{C70} and \Endo{X}{C80}, this ratio is approximately 3. For the longer fullerenes where $B>ZPE$, this ratio decreases to approximately 1.5
		
		Due to the highly regular nature of the wavefunctions of these lowest 50 eigenstates, there is no discernible coupling between the $z$ and $xy$ modes for the low lying energy states, which was also observed in \Endo{Ne}{C70}\cite{mandziukQuantumThreeDimensional1994} and \Endo{H2}{C70}\cite{xuCoupledTranslationrotationEigenstates2009}	. However, a weak coupling between higher energy states of the appropriate symmetry may become apparent, as the energy gap between them decreases and the anharmonic nature of the potential allows them to mix, which has also been seen in \Endo{Ne}{C70}\cite{mandziukQuantumThreeDimensional1994}	.

%\end{document}

%% file: Conclusion.tex
%\documentclass[./X_in_C70]{subfiles}
%\graphicspath{{\subfix{./figures/}}}
%\begin{document}
	We have explored the parameter space of an endohedral atom trapped inside \ce{C70}, interacting via a summed Lennard-Jones potential with the Hamiltonian as given in Equation \eqref{eq:LJ Hamiltonian} and the mass fixed to be the two-particle reduced mass of \Endo{Ne}{C70}. By varying the two LJ parameters, a wide range of symmetric double well potentials were produced, with a variety of different barrier heights $B$, and positions of minima $z_{\text{min}}$. These quantities may also be accessible experimentally, with the latter being seen in diffraction or NMR experiments which could also give information about the localisation of the endohedral atom in the centre or at the minima. The barrier height could possibly be calculated from the tunnelling splitting, which if on the order of 1\wavenumber\, or greater can be visible spectroscopically.
	
	The Hamiltonians constructed from the differing $(\varepsilon, \sigma)$ values were diagonalised using a non-orthogonal basis set, built from 1D harmonic oscillator wavefunctions expanded from the minima. Attempting to use PODVR basis sets built from 1D HO wavefunctions failed when $B$ grew too large as minimising $\alpha$ in Equation \eqref{eq: Variational Principle} leads to a very small effective harmonic potential as the double well weighs down the parabola. Sinc functions also had problems converging to a tight upper bound, especially in the region where $B>ZPE$, seen in Fig \ref{fig: DualMin vs Sinc Convergence}. The HO DVR functions required fixing the interval of sampling points and had an oscillatory behaviour in the convergence. The non-orthogonal basis set seemed better behaved, requiring a similar number of quadrature points to converge as the HO DVR functions and was the basis set of choice. Its possible linear dependence was alleviated by canonically orthogonalising the functions before solving the eigenvalue problem in the subspace. This non-orthogonal basis set converged the ground state energy to milli-wavenumber accuracy. As the non-orthogonal basis set performs similarly to the HO DVR functions when there is a single trapped species, it may be better suited when considering multiple encapsulated species for use in a configuration basis\cite{felkerNuclearorbitalConfigurationinteractionStudy2013}.
		
	The results of the diagonalisation of the three-dimensional Hamiltonian for a variety of $(\varepsilon, \sigma)$ partitioned the space into two regions in two separate ways. One is where the partition lies along the $B=ZPE$ contour, splitting the LJ parameter space into regions where $B>ZPE$ or $B<ZPE$. As the ratio of $\frac{ZPE}{B}$ decreases, the relevance of the non-orthogonal basis set increases as the endohedral species sits deeper in the potential wells. The other partition is where the ground state wavefunction has a single peak, or two maxima where the changeover occurs at $\kappa_z=2.2$. These two differing partitions do not occur in the same place, as there is a smooth deformation of the wavefunction from a single peak which is both squashed and stretched apart to show two maxima well before the $ZPE$ drops down to the value of $B$.

	The wavefunctions were classified based on their statistical moments, namely the standard deviation and kurtosis. The standard deviation gives a measure of the amount of wavefunction density away from the mean, in this case the origin but it cannot distinguish the peakedness of the wavefunction. The kurtosis on the other hand is a good measure of the tailedness of the distribution, describing the density around $\pm\sigma_z$. When the kurtosis drops below 2.2, this is a good indication of when the wavefunction moves away from have a single peak at the origin and starts exhibiting two maxima. These statistics are only part of the covariance and co-kurtosis tensors, which are linked to the quadrupole and hexadecapole moment of the system, which perhaps could be accessible experimentally. The shape of the wavefunction could also be probed using inelastic neutron scattering, which could allow for information of its moments to be calculated.
	
	Excited states of \ce{X}@\ce{C}$_{\mathrm{n}}$	for fullerenes in the $D_{5d/h}$ series from \ce{C70} to \ce{C100} were also calculated, alongside the frequencies of the fundamental transitions of \Endo{X}{C70}. These transitions could be seen spectroscopically and in combination with other diffraction and NMR data, the potential of the endohedral atom could be reconstructed and the best LJ parameters of the \ce{X-C} interaction can be found.  The excited $z$ and $xy$ modes tend to show negative anharmonicity, due to the high degree nature of the potential. Eigenstates lying below the barrier height show a small positive anharmonicity, which persists until the barrier height is exceeded and the negative anharmonicity reappears.

	While the parameter space explored was small, the space can be expanded straightforwardly. In the region where $ZPE<B$, extending the LJ parameter space can introduce a radial double well and we envisage extending our methodology into this space. We aim to use a more accurate potential energy surface, based on some electronic structure calculations which can accurately describe the non covalent interactions\cite{cioslowskiElectronicStructureCalculations2023} in order to converge the eigenstates to a higher accuracy, and compare how well this model LJ potential describes the \ce{X-C} interaction. Varying the LJ parameters was the alternative to changing the encapsulating cage and traversing the $D_{5h/d}$ series\cite{schusslbauerExploringThresholdFullerenes2022}, as $B$ and $z_{\text{min}}$ were more sensitive to this. We intend to compare this non-orthogonal basis set to 1D HO functions when describing the centre of mass of a small molecular species e.g. \ce{H2} by tracking its eigenstates\cite{xuCoupledTranslationrotationEigenstates2009} and ground state wavefunction behaviour along the $D_{5h/d}$ fullerene/fullertube series.

%\end{document}